# Extension of Vertical Equilibrium Model for Two-Phase Displacement in Layer-Cake Reservoirs: Accounting for Water-CO$_2$ Partial Miscibility and Fines Migration.

Kofi Ohemeng Kyei Prempeh[1], Pavel Bedrikovetsky[1], Rouhi Farajzadeh[2].

[1] School of Chemical Engineering, University of Adelaide, Australia
[2] Shell Global Solutions International, The Hague, The Netherlands

## Abstract

Numerical simulations of geological CO$_2$ storage in deep saline aquifers have demonstrated that vertical equilibrium (VE) models offer a robust and computationally efficient framework for reservoir optimization and upscaling. These studies emphasize the influence of fines migration and partial miscibility between water and CO$_2$ on the evolving storage behaviour. Specifically, capillary-driven mobilization of clay and silica fines can impair injectivity, while water evaporation into the CO$_2$ phase leads to near-wellbore drying, salt precipitation, and additional injectivity loss. However, conventional VE models do not account for these coupled physical and geochemical mechanisms. This work introduces an extended VE model that incorporates water vaporization into CO$_2$, CO$_2$ dissolution into the displaced brine, and fines migration leading to permeability reduction. For layer-cake aquifers, the model admits an exact analytical solution, providing closed-form expressions for both injectivity decline and sweep efficiency. The analysis identifies two distinct fronts during displacement: a leading displacement-dissolution front and a trailing full-evaporation front. Vertical model downscaling further reveals that gas saturation varies with the depth-dependent permeability profile. In formations where permeability decreases with depth, saturation declines monotonically; in contrast, in systems with increasing permeability, the competition between viscous and gravitational forces can generate non-monotonic saturation profiles. The results also indicate that while fines migration reduces injectivity, it may simultaneously enhance sweep efficiency; highlighting a complex trade-off that must be considered in the design and optimization of CO$_2$ storage operations.

## Nomenclature
### *Parameters*

| | | |
|---|---|---|
| $A$ | Inverse of pressure diffusivity coefficient, [-] |
| $C_g$ | Overall CO$_2$ component concentration, [-] |
| $\tilde{C}_g$ | CO$_2$ component at the microscale, [-] |
| $F$ | Fractional flow of gas, [-] |
| $f_0$ | Fractional flow of gas in zone 0, [-] |
| $f_{01}$ | Fractional flow of gas in zone 01, [-] |
| $f_1$ | Fractional flow of gas in zone 1, [-] |
| $F$ | Overall fractional flow of CO$_2$ component, [-] |
| $g$ | Gravitational acceleration, [LT$^{-2}$] |
| $H$ | Total height of reservoir, [L] |
| $h_c$ | Dimensionless height of transition zone, [-] |
| $J$ | Leverett J function, [-] |
| $J_w$ | Well Impedance, [-] |
| $j$ | Inverse of Leverett function, [-] |
| $k(z)$ | Horizontal permeability as a function of depth, [L$^2$] |
| $k(Z)$ | Horizontal permeability as a function of dimensionless depth, [L$^2$] |
| $K(Z)$ | Dimensionless horizontal permeability as a function of dimensionless depth, [-] |
| $k_{min}$ | Minimum permeability, [L$^2$] |
| $k_{max}$ | Maximum permeability, [L$^2$] |
| $k_0$ | Average permeability, [L$^2$] |

| | | |
|---|---|---|
| $k_{rg}$ | Gas relative permeability, [-] | |
| $k_{rw}$ | Water relative permeability, [-] | |
| $k_{rgwi}$ | End point gas relative permeability, [-] | |
| $k_{rwgr}$ | End point water relative permeability, [-] | |
| $n_g$ | Corey fluid index for gas, [-] | |
| $n_w$ | Corey fluid index for water, [-] | |
| $p$ | Pressure, $[ML^{-1}T^{-2}]$ | |
| $\tilde{p}$ | Microscale pressure, $[ML^{-1}T^{-2}]$ | |
| $P$ | Dimensionless pressure, [-] | |
| $p_c$ | Capillary pressure $[ML^{-1}T^{-2}]$ | |
| $p_d$ | Capillary pressure in water filled region $[ML^{-1}T^{-2}]$ | |
| $p_e$ | Reservoir pressure, $[ML^{-1}T^{-2}]$ | |
| $p_m$ | Capillary pressure in gas filled region $[ML^{-1}T^{-2}]$ | |
| $P_e$ | Dimensionless reservoir pressure, [-] | |
| $Q$ | Rate, $[L^3T^{-1}]$ | |
| $r$ | Radial coordinate, [L] | |
| $r_w$ | Well radius, [L] | |
| $R$ | Drainage radius, [L] | |
| $S$ | Averaged gas saturation, [-] | |
| $s^*$ | Microscale gas saturation, [-] | |
| $s$ | Normalized microscale gas saturation, [-] | |
| $S_0$ | Averaged gas saturation in zone 0, [-] | |
| $S_{01}$ | Averaged gas saturation in zone 01, [-] | |
| $S_1$ | Averaged gas saturation in zone 1, [-] | |
| $t$ | Time, [T] | |
| $\tau$ | Dimensionless time [-] | |
| $U$ | Total flux, $[LT^{-1}]$ | |
| $U_g$ | Overall $CO_2$ component flux, $[LT^{-1}]$ | |
| $U_{gr}^*$ | Radial gas flux, $[LT^{-1}]$ | |
| $U_{wr}^*$ | Radial water flux, $[LT^{-1}]$ | |
| $U_{wr}$ | Overall radial water flux, $[LT^{-1}]$ | |
| $U_{wz}$ | Overall vertical water flux, $[LT^{-1}]$ | |
| $\tilde{u}_{gr}$ | Radial microscale gas flux, $[LT^{-1}]$ | |
| $\tilde{u}_{gz}$ | Vertical microscale gas flux, $[LT^{-1}]$ | |
| $x$ | Dimensionless length, [-] | |
| $x_w$ | Dimensionless well length, [-] | |
| $z$ | Depth, [L] | |
| $Z$ | Dimensionless depth, [-] | |
| $z_0$ | Position of the advanced front, [L] | |
| $z_1$ | Position of the receded front, [L] | |
| $\Delta P$ | Dimensionless pressure drops, [-] | |

**Greek Symbols**

| | | |
|---|---|---|
| $\beta$ | Formation damage coefficient, [-] | |
| $\phi$ | Porosity, [-] | |
| $\mu$ | Gas-water viscosity ratio, [-] | |
| $\mu_g$ | Gas viscosity, $[ML^{-1}T^{-1}]$ | |
| $\mu_w$ | Water viscosity, $[ML^{-1}T^{-1}]$ | |
| $\rho^G$ | $CO_2$ density, $[ML^{-3}]$ | |
| $\rho_g$ | Gaseous phase density, $[ML^{-3}]$ | |
| $\rho^W$ | | Water density, [ML$^{-3}$] |
| $\rho_w$ | | Aqueous phase density, [ML$^{-3}$] |
| $\kappa$ | | Pressure-diffusivity coefficient |
| $\xi$ | | Self-similar variable, [-] |
| $\sigma_0$ | | Initial particle concentration, [-] |
| $\sigma_s$ | | Strained particle concentration, [-] |
| $\gamma$ | | Dimensionless parameter, [-] |
| $\lambda$ | | Total mobility, [M$^{-1}$L$^3$T] |
| $\lambda_g$ | | Gas mobility, [M$^{-1}$L$^3$T] |
| $\lambda_w$ | | Water mobility, [M$^{-1}$L$^3$T] |
| $\Lambda$ | | Dimensionless total mobility, |
| $v$ | | Relative range for permeability heterogeneity, [-] |
| $\eta$ | | Corey index for Leverett function, [-] |

# 1  Background

Effective site screening for geological $CO_2$ storage, definition of optimal operational parameters, and reliable prediction of $CO_2$ plume migration relies heavily on the results of reservoir simulation. However, the inherent uncertainties in geological and hydrodynamic properties necessitate extensive sensitivity analyses through multi-variant model runs, rendering the simulation process computationally intensive and often impractically time-consuming. To address this challenge, current research efforts have focused on reducing the dimensionality of the flow problem. One promising approach is the use of Vertical Equilibrium (VE) models, which introduce VE pseudo-functions and corresponding upscaling techniques to simplify the simulation framework [1-10]. VE models have proven useful in supporting site selection for geological $CO_2$ storage, as well as in the planning and design of $CO_2$ storage projects in deep saline aquifers and depleted gas fields [11, 12]. Conventional VE models are currently formulated for immiscible displacement scenarios [13]. Variations of the VE immiscible model have been developed under different assumptions, particularly for gravity-dominated and capillary-dominated flow regimes. A representative, though incomplete, selection of the extensive literature on these topics includes the following references [14-27]. The introduction of pseudo-functions in VE models transforms two-dimensional $(x, z, t)$ or $(r, z, t)$ flow problems into simplified one-dimensional $(x, t)$ or $(r, t)$ representations, allowing for explicit analytical solutions. These solutions can be readily used for rapid reservoir prediction and numerical grid upscaling. Moreover, VE models offer reliable substitutes for cumbersome inverse reservoir characterisation problem by analytical Welge-Johnson-Bossler-Naumann (JBN) methods [13, 28].

Evaporation of brine into the injected $CO_2$ and dissolution of $CO_2$ into the displaced brine can significantly influence the performance and dynamics of geological $CO_2$ storage [29-31]. The dissolution of $CO_2$ in formation brine leads to the formation of carbonic acid, which initiates a series of mineral dissolution and precipitation reactions that can alter reservoir properties. Concurrently, water evaporation into the $CO_2$ phase induces near-wellbore drying, triggering salt precipitation and consequent decline in well injectivity [32]. In contrast to fully segregated flow models, incorporating capillary pressure transforms the sharp water-gas interface into a capillary transition zone, which may extend beyond the reservoir thickness. Capillary forces acting on attached reservoir fines at the advancing water-gas menisci can detach and mobilize these particles, resulting in fines migration. This process causes permeability decline and further impairs injectivity, while also influencing overall sweep efficiency [33-40]. Despite the significance of these coupled processes, current VE models for $CO_2$-brine displacement remain limited to immiscible formulations and do not capture the effects of partial miscibility, capillarity, or fines migration.

Fig. 1 illustrates the schematic of radial water displacement by gas in a layer-cake reservoir, where permeability $k(z)$ decreases with depth. The mixed water-gas zone propagates behind the advancing front of formation brine. The gas saturation $s=0$ at the water-gas front, denoted $z_0(r,t)$, marks the leading edge of the displacement. The zone of dry, injected $CO_2$ follows the mixed zone, with gas saturation $s=1$ at the evaporation front $z_1(r,t)$. Fig. 3b depicts the case of a thin reservoir, where the capillary transition zone thickness exceeds the reservoir thickness, i.e., $z_1(r,t) - z_0(r,t)>H$. In contrast, Fig. 3a shows a thick reservoir, where the reservoir thickness exceeds that of the transition zone, $z_1(r,t) - z_0(r,t)<H$.

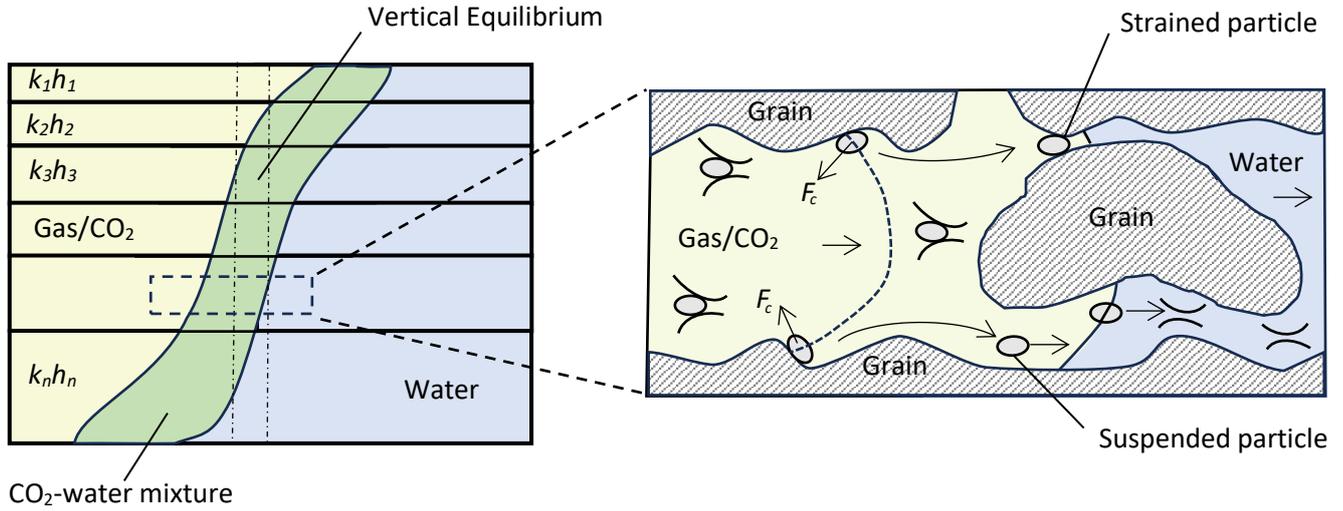

*Fig. 1: Schematic for two-phase flow with fines migration (fines detachment, migration and straining) in porous media.*

The Vertical Equilibrium (VE) model is valid when viscous pressure gradients are negligible compared to capillary and gravitational forces. In cases where the reservoir thickness significantly exceeds that of the capillary transition zone, the VE model degenerates to a segregated flow regime, with gas flowing above water in distinct layers [28]. Conversely, when the transition zone is much thinner than the reservoir, the system approaches a capillary-dominated two-phase flow scenario [41, 42]. Under these conditions, the assumption that the displacing phase sequentially invades layers from highest to lowest permeability gives rise to the classical Dietz pseudo-functions [13, 41-44]. However, in more general cases where viscous forces are non-negligible, this restrictive assumption no longer holds. More advanced flow models for viscous-dominant displacement waive the restrictive assumption of the filling of the layers by the displacing phase in order of permeability decrease [24, 25, 45, 46].

We consider a water-wet porous medium, where capillary pressure increases with gas saturation $s$ (Fig. 2b), ranging from the entry pressure $p_d$ at full water saturation to a maximum value $p_m$ in the absence of water. These capillary pressures – $p_d$ and $p_m$ – correspond to the advanced and receded displacement fronts, respectively, as shown in Fig. 3. Partial miscibility between water and $CO_2$ is represented by the mole fraction of $CO_2$ dissolved in water ($c_g$) and the mole fraction of water evaporated into the injected $CO_2$ ($c_w$) (Fig. 2a). These quantities describe the mutual solubility of the two phases per unit pore volume and are functions of pressure and temperature under thermodynamic equilibrium conditions (Fig. 2c). At the pore scale, these mass transfer processes occur across capillary-controlled water-gas menisci, which advance through the pore space during displacement. Water evaporation into the injected $CO_2$ phase leads to salt precipitation from the brine, resulting in near wellbore plugging and a significant reduction in injectivity [29, 46].

Current VE models for immiscible flow reduce the two-dimensional (2D) displacement problem to an exact Buckley-Leverett solution for a "pseudorized" one-dimensional (1D) flow domain [13]. Similarly, the 1D two-phase displacement problem incorporating partial miscibility also admits an exact solution. Like the classical Buckley–Leverett formulation, this solution includes shock fronts and allows for graphical interpretation [28]. Fig. 4a presents the phase diagram for a fixed pressure / temperature, corresponding to the thermodynamic conditions illustrated in Fig. 2c. The flux of $CO_2$ in both phases depends on the overall $CO_2$ concentration, $C_g$, within the pore space. This functional relationship, $F(C_g)$, is shown as a continuous curve in Fig. 4b. The endpoints of the curve correspond to the displaced (pure water, $C_g=0$) and injected (pure $CO_2$, $C_g=1$) states. The resulting concentration profile of $CO_2$ is shown in Fig. 4c. Starting from the injection point, the displacement domain consists of the following zones: (1) a dry $CO_2$ injection zone with $C_g=1$; (2) a full evaporation front (transition from point 1 → 2); (3) a mixed-phase flow zone (2 →3); (4) a water–gas front (transition from point 3 →1); and (5) the flow zone of the undisturbed reservoir brine.

Fines migration is a well-documented phenomenon in $CO_2$ storage projects [33-40]. Naturally occurring reservoir fines, such as clays and silica particles, are typically adhered to rock surfaces and may be detached by drag, lift, or capillary forces. While these particles often form surface coatings and their detachment does not noticeably enhance permeability, their subsequent migration can lead to severe flow impairment. In particular, straining (size exclusion) of mobilized fines in narrow pore throats can drastically alter local flow dynamics, potentially causing a complete loss of permeability and hydraulic connectivity [47]. The graphical abstract illustrates this process at the pore scale as a magnified view of the reservoir-scale system. Fines are detached by capillary forces and transported along advancing gas-water menisci, ultimately becoming trapped in constricted pores. Displacement proceeds in a piston-like fashion, with sequential meniscus positions indicated by dashed and solid lines. Residual, capillary-trapped water in micro-pores is omitted for clarity. Strained or re-attached fines remain within the gas-occupied pore space, reducing permeability to the displacing $CO_2$ phase while having negligible effect on the already displaced water.

Despite the significant impact of fines migration, partial $CO_2$-water miscibility, and capillary pressure on $CO_2$ injection into saline aquifers, existing Vertical Equilibrium (VE) models do not incorporate these coupled physical processes. This study addresses this gap by developing an extended VE framework that accounts for these critical phenomena.

The quasi-2D governing system for gas–water flow under Vertical Equilibrium (VE) conditions, accounting for mutual $CO_2$-water miscibility and fines migration, is derived. This formulation reduces to 1D hyperbolic conservation law that admits an exact self-similar solution for the displacement of water by $CO_2$. The solution includes explicit formulae for phase saturations, component concentrations, and reservoir pressure. To match the pressure distribution in the incompressible region behind the displacement front with the initial reservoir pressure, a small compressibility is introduced ahead of the displacement front. The VE assumption enables downscaling of the analytical 1D solution in radial–temporal coordinates ($r,t$) to reconstruct full vertical distributions ($r,z,t$) of saturation, concentration, and pressure. The downscaled profiles reveal that in formations where permeability decreases with depth, gas saturation decreases monotonically with depth. Conversely, in formations with increasing permeability with depth, the competition between viscous and gravitational forces may produce non-monotonic vertical saturation profiles. The model further demonstrates that fines migration and the resulting permeability reduction lead to a decline in well injectivity, while enhancing reservoir sweep and increasing overall $CO_2$ storage capacity. Sensitivity analysis based on the analytical solution identifies the most influential parameters as the formation damage coefficient, the concentration of mobile fines, and the reservoir heterogeneity index.

The structure of the manuscript is as follows. Section 1 introduces the key physical phenomena influencing water displacement by $CO_2$, which are incorporated into the proposed model. Section 2 outlines the model assumptions and derives the governing equations for both incompressible and compressible zones. Section 3 presents vertically averaged "pseudo-function" model and pressure diffusivity equation in undisturbed zone. Section 4 derives the self-similar solution to the averaged model and analytical formulae for pressure in the low compressible undisturbed zone as well as total mobility. Section 5 presents explicit formulae for well injectivity and sweep efficiency calculations. Section 6 details the downscaling procedure used to reconstruct full vertical profiles from the 1D solution. Section 7 conducts a sensitivity analysis with respect to reservoir properties and injectivity parameters. Sections 8 and 9 provide a discussion of results and concluding remarks.

## 2 Mathematical model

This section presents the main assumptions of VE model for displacement in layer-cake reservoirs accounting for partial $CO_2$-brine miscibility and fines migration (section 2.1) and schematic for flow zone structure (section 2.2).

### 2.1 Basic assumptions of new VE model

*Flow zones with water-gas mixture* – The aqueous and gaseous phases are assumed to contain water and $CO_2$ components. The mixing of these components is described using Amagat's law of additive volumes, under the assumption that the individual densities of $CO_2$ and water remain constant and equal in both phases. This allows for the introduction of volumetric concentrations for each component in the respective phases.

$$\rho_g = c_w \rho^W + (1-c_w)\rho^G, \quad \rho_w = c_g \rho^G + (1-c_g)\rho^W \tag{1}$$

where $\rho^W$ and $\rho^G$ correspond to densities of pure components, while $\rho_g$ and $\rho_w$ are phase densities. Pressures and temperatures are constant. Phase equilibrium is assumed, so $c_w$ and $c_g$ are also constant, representing the fraction of water vaporised into gas and the fraction of gas dissolved in water, respectively, as shown in Fig. 2a.

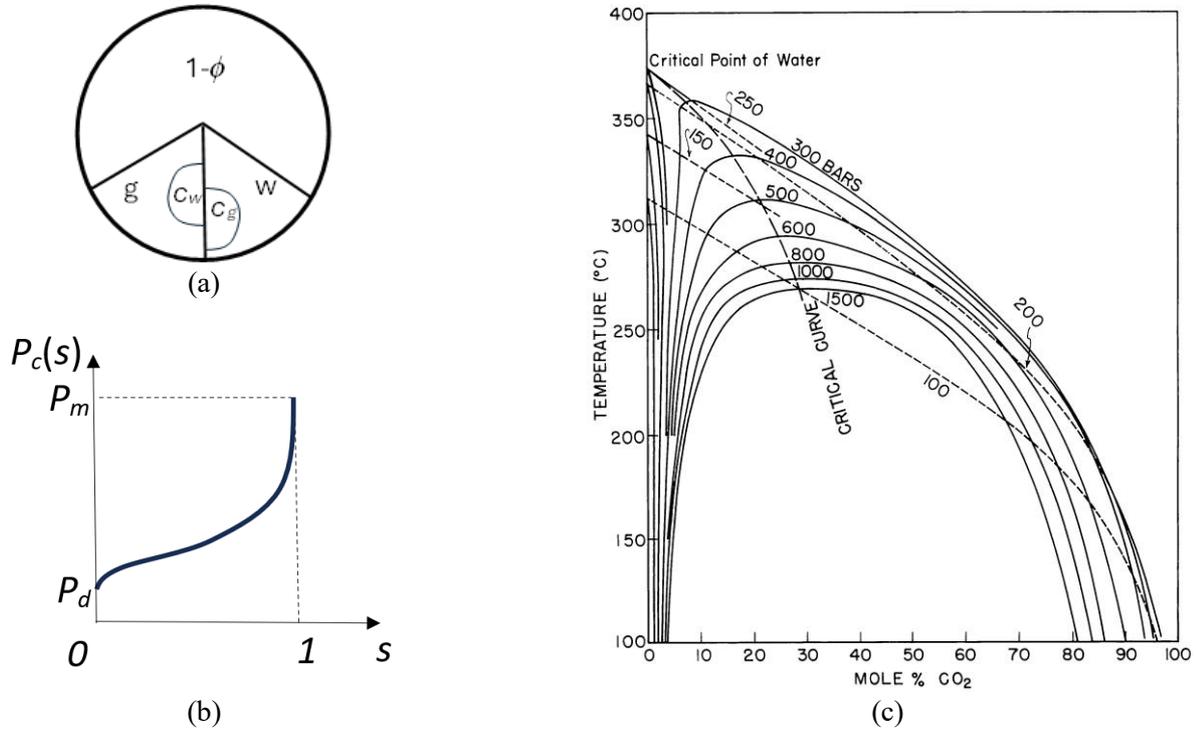

Fig. 2: Schematic for binary water-CO$_2$ mixture in porous media: (*a*) volumetric fractions of CO$_2$ and water components in pore volume; (*b*) Capillary pressure curve as a function of gas saturation, s; (*c*) isobaric phase diagram for binary water-CO$_2$ system.

The volumetric CO$_2$ concentration and flux of gas are given as

$$\tilde{C}_g = (1-c_w)s + c_g(1-s), \quad \tilde{u}_{gr} = (1-c_w)u_{gr} + c_g u_{wr} \tag{2}$$

Hydrostatic pressure is assumed for each phase, i.e.

$$\frac{\partial p_g}{\partial z} = -\rho_g g, \quad \frac{\partial p_w}{\partial z} = -\rho_w g \tag{3}$$

The rock remains water-wet during the overall displacement

$$p_c(s) = p_g - p_w \tag{4}$$

where the capillary pressure is expressed via dimensionless Leverett function $J(s)$

$$p_c(s) = \frac{\sigma \cos\theta}{\sqrt{k(z)/\phi}} J(s) \tag{5}$$

The model accounts for formation damage induced by fines migration. It is assumed that at the reservoir scale, fines detachment, migration and straining occur instantly if compared with the time in *1* Pore-Volume-Injected (PVI). The detaching capillary force, exerting on the attached particle, is significantly higher than attaching electrostatic force. So, all detachable fines are detached by the advanced gas-water menisci during passing via a pore. Therefore, the concentration of attached fines in the rock is a function of saturation [48, 49]. It is called the maximum retention function (MRF). Lab tests show that MRF can be approximated by power-law dependency [49]

$$\sigma_{cr}(s) = \sigma_1(s)^B \tag{6}$$

The concentration of strained fines at saturation *s* is equal to the difference of the initial and current attached concentrations:

$$\Delta\sigma_{cr} = \sigma_{cr}(1) - \sigma_{cr}(s), \tag{7}$$

Modification to Darcy's law for two-phases gives the expressions for phase velocities accounting for the formation damage factor in the gas phase only.

$$u_{wr} = -\frac{k(z)k_{rw}(s)}{\mu_w}\frac{\partial p_w}{\partial r}, \quad u_{gr} = -\frac{k(z)k_{rg}(s)}{\mu_g\left[1+\beta\Delta\sigma_{cr}(s)\right]}\frac{\partial p_g}{\partial r} \tag{8}$$

The schematic in the graphical abstract illustrates the behaviour of detached fines that migrate through the porous medium, transported by the detaching meniscus, until they are retained by straining. As shown, a migrating particle may become trapped within a narrow pore throat, effectively plugging the pore. Once straining occurs, the ensemble of menisci continues to advance beyond the obstructed region. Consequently, the presence of strained particles results in a reduction of permeability within the flow domain of the displacing phase, while the displaced phase remains unaffected. Therefore, in Eq. (8), the damage is expressed exclusively for the gas phase.

*Undisturbed zone of displaced reservoir fluid flow* – In the single-phase zone containing the displaced reservoir fluid, the hydrostatic pressure distribution is described by the second equation in Eq. (3). In this region, both the reservoir brine and the rock matrix are assumed to exhibit low compressibility, i.e.,

$$\phi(p)\rho_w(p) = \phi_0 \rho_{w0} \exp\left[-\frac{p - p_0}{c_t}\right] \tag{9}$$

where index "*0*" corresponds to "initial conditions", and $c_t$ is the total compressibility coefficient for fluid and rock [50].

*Dry zone of the displacing fluid flow* – Here, the depth hydrostatic pressure distribution is assumed too. Yet, the displacing gas is assumed to be incompressible. Index "*1*" corresponds to the gas zone in Fig. 3a, b.

## 2.2 Schematic of displacement zone

Fig. 3 illustrates the schematic of water displacement by $CO_2$ in a layer-cake reservoir. Ahead of the displacement front, at $z<z_0(r,t)$, no $CO_2$ is present; thus, the gas saturation is $s=0$ and the $CO_2$ concentration is $c_g=0$ in this region. Conversely, behind the rear front of complete evaporation, at $z>z_1(r,t)$, no water remains, resulting in zone $s=1$ and $c_g=1$. Therefore, the saturation is $s=0$ ahead of the displacement front and $s=1$ behind the rear front.

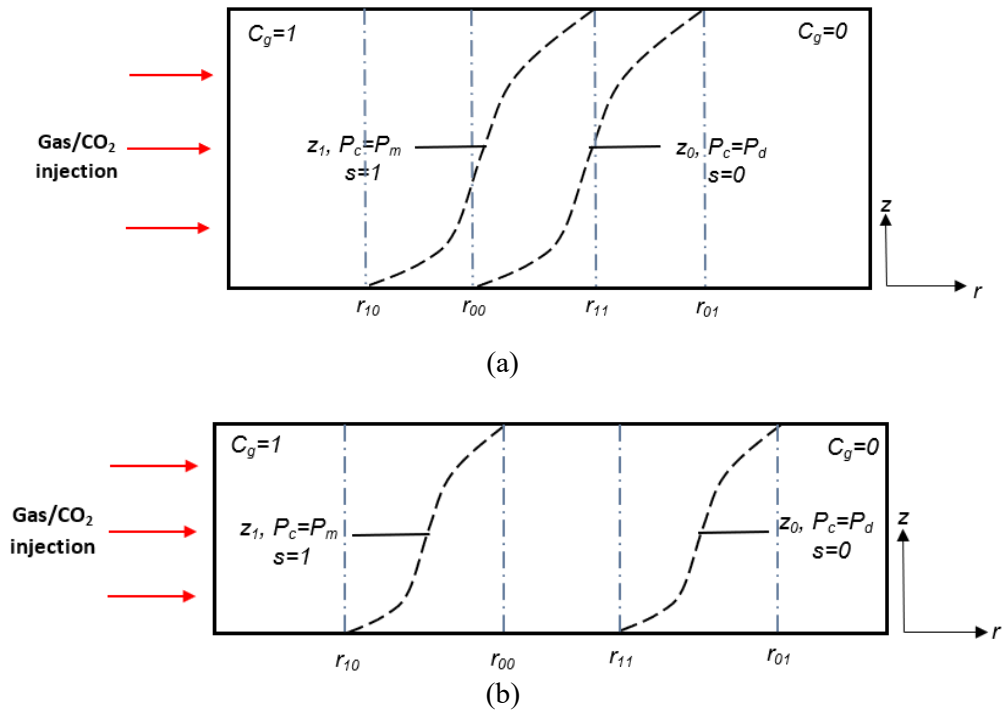

*Fig. 3: Schematic for water displacement by $CO_2$ accounting for partial miscibility of phases: (a) in thick reservoir; (b) in thin reservoir.*

## 3 Homogenisation of 2D model and "pseudo-functions"

This section derives the averaged flow equation – VE model – including development of the fractional flow expressions (section 3.1), derivation of volumetric balance equations (section 3.2), and obtaining the solution for pressure in the flow zone of low-compressible fluid (section 3.3).

## 3.1 Calculation for averaged fractional flow

The difference of the hydrostatic pressures given by Eq. (3) yields the first order derivative for capillary pressure as

$$\frac{dp_c}{dz} = \Delta\rho g, \quad \Delta\rho = \rho_w - \rho_g \tag{10}$$

The gas saturation and capillary pressure at the advanced front, $z_0$ and the receded front, $z_1$ are given as:

$$z = z_0(x,t): \quad s = 0, \quad p_c = p_d \tag{11}$$

$$z = z_1(x,t): \quad s = 1, \quad p_c = p_m \tag{12}$$

Taking the integral of Eq. (10) yields

$$p_c(z) = \Delta\rho g z + const \tag{13}$$

Substituting Eqs. (11-12) into Eq. (13) and taking the difference of the resulting equations yields the expression for the thickness of the capillary transition zone, $H_c$.

$$p_m - p_d = \Delta\rho g H_c, \quad H_c = z_1 - z_0 \tag{14}$$

*Thick Reservoirs ($H > H_c$), Fig. 3a.* – We now derive averaged equations for the case of thick reservoirs, where the reservoir thickness, $H$, exceeds the thickness of the transition zone, $H_c$.

Taking the integral of Eq. (10) from $z$ to $z_0$, accounting for Eq. (11) yields the expression for capillary pressure at the advanced front as

$$p_c(s) - p_d = \Delta\rho g(z - z_0) = \frac{\sigma[J(s) - J_d]}{\sqrt{k(z)/\phi}}, \quad J(s) = J_d + (z - z_0)\frac{\Delta\rho g}{\sigma}\sqrt{k(z)/\phi} \tag{15}$$

Considering the inverse function to Eq. (15), $j = J^{-1}$, the vertical saturation distribution is obtained as

$$s = j\left(J_d + (z - z_0)\frac{\Delta\rho g}{\sigma}\sqrt{k(z)/\phi}\right) \tag{16}$$

Substituting Eq. (16) into the first equation of Eq. (2) gives gas concentration in two-phase zone at $z=z_0$ as

$$\tilde{C}_g = j\left(J_d + (z - z_0)\frac{\Delta\rho g}{\sigma}\sqrt{k(z)/\phi}\right)(1 - c_w - c_g) + c_g \tag{17}$$

Similarly, taking the integral of Eq. (10) from $z$ to $z_1$, accounting for Eq. (12) yields the expression for capillary pressure at the receded front as

$$p_c(s) - p_m = \Delta\rho g(z - z_1) = \frac{\sigma[J(s) - J_m]}{\sqrt{k(z)/\phi}}, \quad J(s) = J_m + (z - z_1)\frac{\Delta\rho g}{\sigma}\sqrt{k(z)/\phi} \tag{18}$$

Substituting the inverse function of Eq. (18), $j = J^{-1}$ into Eq. (2) gives the gas concentration in two-phase zone at $z=z_1$ as

$$\tilde{C}_g = j\left(J_m + (z - z_1)\frac{\Delta\rho g}{\sigma}\sqrt{k(z)/\phi}\right)(1 - c_w - c_g) + c_g \tag{19}$$

The average gas saturation across the depth of the reservoir is determined by the expression

$$S(r,t) = \frac{1}{H}\int_0^H s(r,z,t)dz \tag{20}$$

where Zone 0, Zone 01 and Zone 1 represent the region with "mixture above water", "mixture between gas and water" and "mixture below gas" respectively as illustrated in Fig. 3a, b; where,

$$0: r_{11} < r < r_{01}, \qquad 01: r_{00} < r < r_{11}, \qquad 1: r_{10} < r < r_{00} \tag{21}$$

The fractional flow of gaseous phase is defined in terms of the phase mobilities, Eq. (A7) as

$$f = \frac{\int_0^H \frac{k(z)k_{rg}(s)dz}{\mu_g[1+\beta\Delta\sigma_{cr}(s)]}}{\left[\int_0^H \frac{k(z)k_{rw}(s)dz}{\mu_w} + \int_0^H \frac{k(z)k_{rg}(s)dz}{\mu_g[1+\beta\Delta\sigma_{cr}(s)]}\right]} = \frac{\lambda_g(s)}{[\lambda_w(s)+\lambda_g(s)]} = \frac{\lambda_g(s)}{\lambda(s)}, \quad (22)$$

where the overall fractional flow of $CO_2$ component is expressed in terms of Eq. (22) as

$$F = (1-c_w)f + (1-f)c_g \quad (23)$$

From Eq. (A6), the overall gas flux (detailed derivation is provided in Appendix A) in terms of the overall fractional flow of $CO_2$ (Eq. (23)) is obtained as

$$U_g = UF + [F-(1-c_w)]\int_0^H \frac{k(z)k_{rg}(s)dz}{\mu_g[1+\beta\Delta\sigma_{cr}(s)]} \frac{\partial p_c(s)}{\partial r} \quad (24)$$

Calculations of overall $CO_2$-component concentration $C_g$ and fractional flow $F$ in thick reservoir are presented in Table 1. The equations for overall gas saturation $S_0$, $CO_2$ concentration $C_0$ and fractional flow for gas phase $f_0(S)$ for the zone 0 "mixture above water" are given in second line of Table 1. Third line of Table 1 contains those three expressions for the zone 01 "mixture between gas and water". The formulae for the zone 1 "mixture below gas" are presented in fourth line. Overall fractional flows for $CO_2$-component $F_0(C)$, $F_{01}(C)$, and $F_1(C_g)$ are obtained using Eq. (23).

*Table 1: Averaged gas component concentration and fractional flow across distinct regions of a thick reservoir.*

| Region | $S(r,t)$ | Overall $CO_2$ concentration $C_g = S(1-c_w)+(1-S)c_g$ | Fractional flow for gas phase $f(S)$ |
|---|---|---|---|
| 0 | $\frac{1}{H}\int_{z_0}^{H} j_0\left(J_d+(z-z_0)\frac{\Delta\rho g}{\sigma}\sqrt{k(z)/\phi}\right)dz$ | $c_g + \frac{(1-c_w-c_g)}{H} \cdot \int_{z_0}^{H} j_0\left(J_d+(z-z_0)\frac{\Delta\rho g}{\sigma}\sqrt{k(z)/\phi}\right)dz$ | $\frac{\int_{z_0}^{H}\frac{k(z)k_{rg}(s)dz}{[1+\beta\Delta\sigma_{cr}(s)]}}{\left[\frac{\mu_g}{\mu_w}\int_0^{z_0}k(z)dz+\int_{z_0}^{H}\frac{k(z)k_{rg}(s)dz}{[1+\beta\Delta\sigma_{cr}(s)]}\right]}$ |
| 01 | $\frac{1}{H}\int_{z_0}^{z_0+H_c} s(z)dz + \frac{(H-H_c-z_0)}{H}$ | $c_g+(1-c_w-c_g)\cdot\left[\frac{1}{H}\int_{z_0}^{z_0+H_c} s(z)dz+\frac{(H-H_c-z_0)}{H}\right]$ | $\left[\int_{z_0}^{z_0+h_c}\frac{k(z)k_{rg}(s)dz}{[1+\beta\Delta\sigma_{cr}(s)]}+\frac{k_{rgwc}}{[1+\beta\Delta\sigma_{cr}(1-S_{wc})]}\int_{z_0+h_c}^{1}k(z)dz\right]\cdot$ $\left\{\frac{\mu_g}{\mu_w}\left(\int_0^{z_0}k(z)dz+\int_{z_0}^{z_0+h_c}\frac{k_{rw}(s)k(z)dz}{\mu_w}\right)+\right.$ $\left.\int_{z_0}^{z_0+h_c}\frac{k(z)k_{rg}(s)dz}{[1+\beta\Delta\sigma_{cr}(s)]}+\frac{k_{rgwc}}{[1+\beta\Delta\sigma_{cr}]}\int_{z_0+h_c}^{1}k(z)dz\right\}^{-1}$ |
| 1 | $\frac{1}{H}\int_0^{z_1} j_1\left(J_m+(z-z_1)\frac{\Delta\rho g}{\sigma}\sqrt{k(z)/\phi}\right)dz + \frac{(H-z_1)}{H}$ | $c_g+(1-c_w-c_g)\cdot\left\{\frac{1}{H}\int_0^{z_1} j_1\left(J_m+(z-z_1)\frac{\Delta\rho g}{\sigma}\sqrt{k(z)/\phi}\right)dz+\frac{(H-z_1)}{H}\right\}$ | $\left[\int_0^{z_1}\frac{k(z)k_{rg}(s)dz}{\mu_g[1+\beta\Delta\sigma_{cr}(s)]}+\frac{k_{rgwc}}{[1+\beta\Delta\sigma_{cr}(1-S_{wc})]}\int_{z_1}^{1}k(z)dz\right]\cdot$ $\left\{\frac{\mu_g}{\mu_w}\int_0^{z_1}k(z)k_{rw}(s)dz+\int_0^{z_1}\frac{k(z)k_{rg}(s)dz}{[1+\beta\Delta\sigma_{cr}(s)]}+\right.$ $\left.\frac{k_{rgwc}}{[1+\beta\Delta\sigma_{cr}]}\int_{z_1}^{1}k(z)dz\right\}^{-1}$ |

*Thin Reservoirs ($H<H_c$), Fig. 3b.* – For the case of thin reservoirs, where the capillary transition zone is thicker than the reservoir, the expression for capillary pressure, vertical saturation distribution, the gas concentration in two-phase zone at the advanced front and receded front are the same as those derived for thick reservoir given by Eqs (15-19). Table 2 contains the equations for overall gas saturation $S(r,t)$ (column two), $CO_2$ concentration (column 3), and fractional flow for gas phase $f(S)$ (column four). Line two presents those formulae for the zone 0 "mixture above water". Those expressions for the zone 01 "mixture between gas and water" are given in line three. The fourth line corresponds to the zone 1 "mixture below gas". The overall fractional flow of $CO_2$-component versus its concentration $F(C_g)$ is calculated using Eq. (23).

*Table 2: Averaged gas component concentration and fractional flow across distinct regions of a thin reservoir.*

| Region | $S(r,t)$ | Overall $CO_2$ concentration $C_g = S(1-c_w) + (1-S)c_g$ | Fractional flow for gas phase $f(S)$ |
|---|---|---|---|
| 0 | $\frac{1}{H}\int_{z_0}^{H} j_0\left(J_d + (z-z_0)\frac{\Delta\rho g}{\sigma}\sqrt{k(z)/\phi}\right)dz$ | $c_g + \frac{(1-c_w-c_g)}{H} \bullet \int_{z_0}^{H} j_0\left(J_d + (z-z_0)\frac{\Delta\rho g}{\sigma}\sqrt{k(z)/\phi}\right)dz$ | $\dfrac{\int_{z_0}^{1} \dfrac{k(z)k_{rg}(s)dz}{[1+\beta\Delta\sigma_{cr}(s)]}}{\left[\dfrac{\mu_g}{\mu_w}\int_{0}^{z_0} K(Z)dZ + \int_{z_0}^{1}\dfrac{k(z)k_{rg}(s)dz}{[1+\beta\Delta\sigma_{cr}(s)]}\right]}$ |
| 01 | $\dfrac{1}{H}\int_{0}^{H} s(r,t)dz$ | $c_g + (1-c_w-c_g)\dfrac{1}{H}\int_{0}^{H} s(r,t)dz$ | $\dfrac{\int_{0}^{H_\varepsilon} \dfrac{k(z)k_{rg}(s)}{[1+\beta\Delta\sigma_{cr}(s)]}dz}{\left[\dfrac{\mu_g}{\mu_w}\int_{0}^{H_\varepsilon}\dfrac{k(z)k_{rw}(s)}{\mu_w}dz + \int_{0}^{H_\varepsilon}\dfrac{k(z)k_{rg}(s)}{[1+\beta\Delta\sigma_{cr}(s)]}dz\right]}$ |
| 1 | $\dfrac{1}{H}\int_{0}^{z_1} j_1\left(J_m + (z-z_1)\dfrac{\Delta\rho g}{\sigma}\sqrt{k(z)/\phi}\right)dz + \dfrac{(H-z_1)}{H}$ | $c_g + (1-c_w-c_g)\bullet \left\{\dfrac{1}{H}\int_{0}^{z_1} j_1\left(J_m + (z-z_1)\dfrac{\Delta\rho g}{\sigma}\sqrt{k(z)/\phi}\right)dz + \dfrac{(H-z_1)}{H}\right\}$ | $\left[\int_{0}^{z_1}\dfrac{k(z)k_{rg}(s)dz}{\mu_g[1+\beta\Delta\sigma_{cr}(s)]} + \dfrac{k_{rgwc}}{[1+\beta\Delta\sigma_{cr}(1-S_{wc})]}\int_{z_1}^{1}k(z)dz\right] \bullet \left\{\dfrac{\mu_g}{\mu_w}\int_{0}^{z_1} k(z)k_{rw}(s)dz + \int_{0}^{z_1}\dfrac{k(z)k_{rg}(s)dz}{[1+\beta\Delta\sigma_{cr}(s)]} + \dfrac{k_{rgwc}}{[1+\beta\Delta\sigma_{cr}]}\int_{z_1}^{1}k(z)dz\right\}^{-1}$ |

## 3.2 Derivation of 1D homogenised continuity equation for gas

The equation for 1D two-phase flow of partly miscible fluids in heterogeneous non-deformable porous media follows from substituting the dimensionless variables and parameters Eq. (B3) into Eq.(B2) to obtain:

$$\frac{\partial C_g}{\partial \tau} + \frac{\partial F}{\partial x} = -\delta \frac{\partial}{\partial x}\left[(F-(1-c_w))\int_{0}^{1}\frac{K(Z)k_{rg}(s)dZ}{\mu(1+\beta\Delta\sigma_{cr}(s))} \times \frac{\partial J(s)}{\partial x}\right] \quad (25)$$

where the expressions for overall fractional flow of $CO_2$ are presented in Table 1 and Table 2.

Large scale approximation corresponds to negligibly small value of parameter where

$$\delta = \frac{4\pi k_0 H}{Q\mu_w}\frac{\sigma\cos\theta\sqrt{\phi}}{\sqrt{k_0}} \ll 1 \quad (26)$$

Considering the following typical values for interfacial tension, $\sigma = 20 mN/m$; contact angle, $\theta = 30^0$; reservoir thickness, $H = 15m$; average absolute permeability, $k_o = 3.4641\text{e-}13 m^2$; viscosity of water, $\mu_w = 1.002 mPas$ and injection rate, $Q = 1.25\text{e-}2\ m^3/s$, the value of $\delta \approx 0.0767 \ll 1$.

For large scale approximations, the mass balance equation Eq. (25) yields a hyperbolic equation given as

$$\frac{\partial C_g}{\partial \tau} + \frac{\partial F(C_g)}{\partial x} = 0 \quad (27)$$

An initial-boundary value problem for Eq (27), describing displacement of water by gas is formulated as follows:

$$\begin{cases} t = 0: & C_g = 0 \\ r = r_w: & C_g = 1 \end{cases} \quad (28)$$

Substitution of the dimensionless parameters Eq. (A3) expresses the initial-boundary value problem as

$$\begin{cases} \tau = 0: \ C_g = 0 \\ x = x_w = \left(\dfrac{r_w}{R}\right)^2 : \ C_g = 1 \end{cases} \quad (29)$$

## 3.3 Derivation of pressure diffusivity equation ahead of water-gas front

Under the assumptions of hydrostatic pressure distribution and low compressibility of water and rock, Appendix C derives the pressure diffusivity equation

$$\frac{\partial P}{\partial \tau} = \kappa \frac{\partial}{\partial x}\left(x \frac{\partial P}{\partial x}\right) \quad (30)$$

where $P$ is the averaged over depth reservoir pressure, $\kappa$ is the pressure-diffusivity coefficient that corresponds to averaged permeability, introduced in Eq. (C5).

## 4 Derivation of the model in large-scale approximation and analytical solution

This section derives the analytical model – solution for averaged VE equations – that includes the expressions for saturations and concentrations in the incompressible flow zone (section 4.1), the solution in the compressible flow zone (section 4.2), and the expression for total two-phase mobility (section 4.3).

### 4.1 Self-similar solution

Let us derive exact solution of initial-boundary problem (27) for Eq. (29). Applying the chain rule to the second term of Eq. (27) yields

$$\frac{\partial C_g}{\partial \tau} + \frac{dF}{dC_g}\frac{\partial C_g}{\partial x} = 0 \quad (31)$$

Introducing the self-similar variable, $\xi = x/t$,

$$\xi = \frac{x}{\tau}, \quad \begin{cases} \partial x = \tau \partial \xi \\ \partial \tau = -\dfrac{\tau}{\xi}\partial \xi \end{cases} \quad (32)$$

The solution to the initial-boundary value problem Eqs. (27, 29) is

$$C_g(\xi) = \begin{cases} 1, & \dfrac{x_w}{\tau} < \xi < D_1 \\ \xi = \dfrac{x}{\tau} = F'(C_g), & D_1 < \xi < D_0 \\ 0, & D_0 < \xi < \infty \end{cases} \quad (33)$$

Mass balance (Hugoniot-Rankine) conditions are fulfilled on the shock fronts $z_0$ and $z_1$. The shock velocities at the advanced and receded front are

$$\frac{dx}{d\tau}(z_0) = D_0 = \frac{F(C_g^+) - F(C_g^-)}{C_g^+ - C_g^-} = \frac{F(C_g^-)}{C_g^-}, \quad \frac{dx}{d\tau}(z_1) = D_1 = \frac{F(C_g^+) - F(C_g^-)}{C_g^+ - C_g^-} = \frac{F(C_g^+) - 1}{C_g^+ - 1} \quad (34)$$

Fig. 4(a-c) illustrates a typical form of fractional flow curve obtained for a binary mixture of water and $CO_2$ that accounts for partial miscibility between water and $CO_2$, and the nature of the self-similar solution in terms of $\xi$

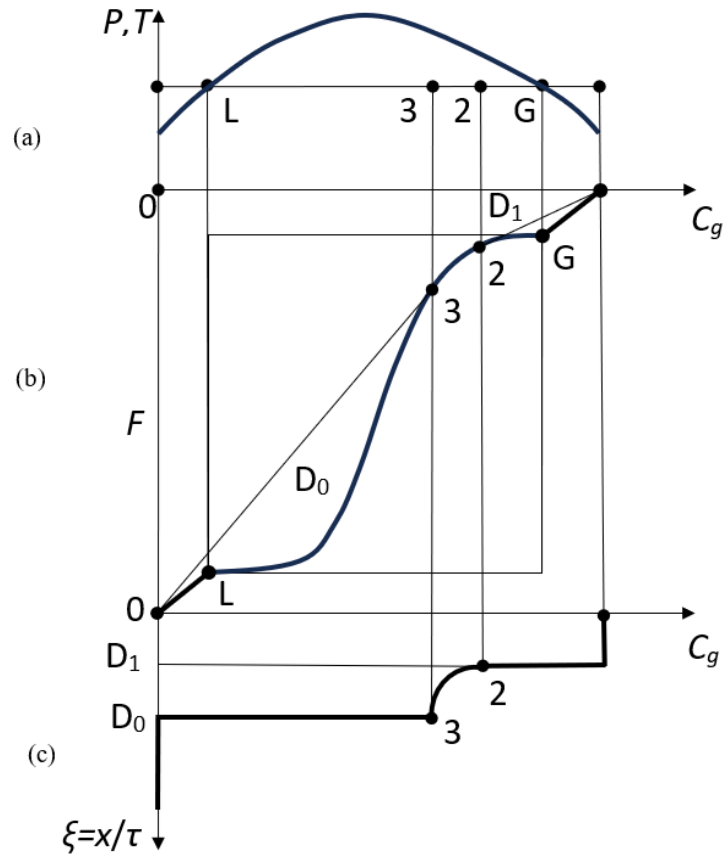

Fig. 4: Graphical representation of fractional flow curve accounting for partial miscibility water-$CO_2$ phases: (a) phase diagram for overall $CO_2$ content in gaseous and aqueous phases; (b) fractional flow versus $CO_2$ concentration; (c) Self-similar profile for $CO_2$ propagation in terms of $\xi$

Fig. 5a presents the plot of fractional flow function $F(C_g)$ for a viscosity ratio of $\mu=\mu_g/\mu_w=0.02$ (blue curve). The plot consists of three distinct regions: a straight-line connecting point *1* and *G*, representing undersaturated single-phase gas flow; an S-shaped curve between points *G* and *L*, corresponding to the two-phase flow region; and a straight-line from point *0* to *L*, denoting undersaturated single-phase water flow as illustrated in Fig. 4b. The slopes of both straight-line segments are unity, indicating that, in the single-phase regions, each component is transported entirely by the total flux of the respective carrier phase. Notably, the fractional flow functions in the single-phase regions are independent of phase viscosities.

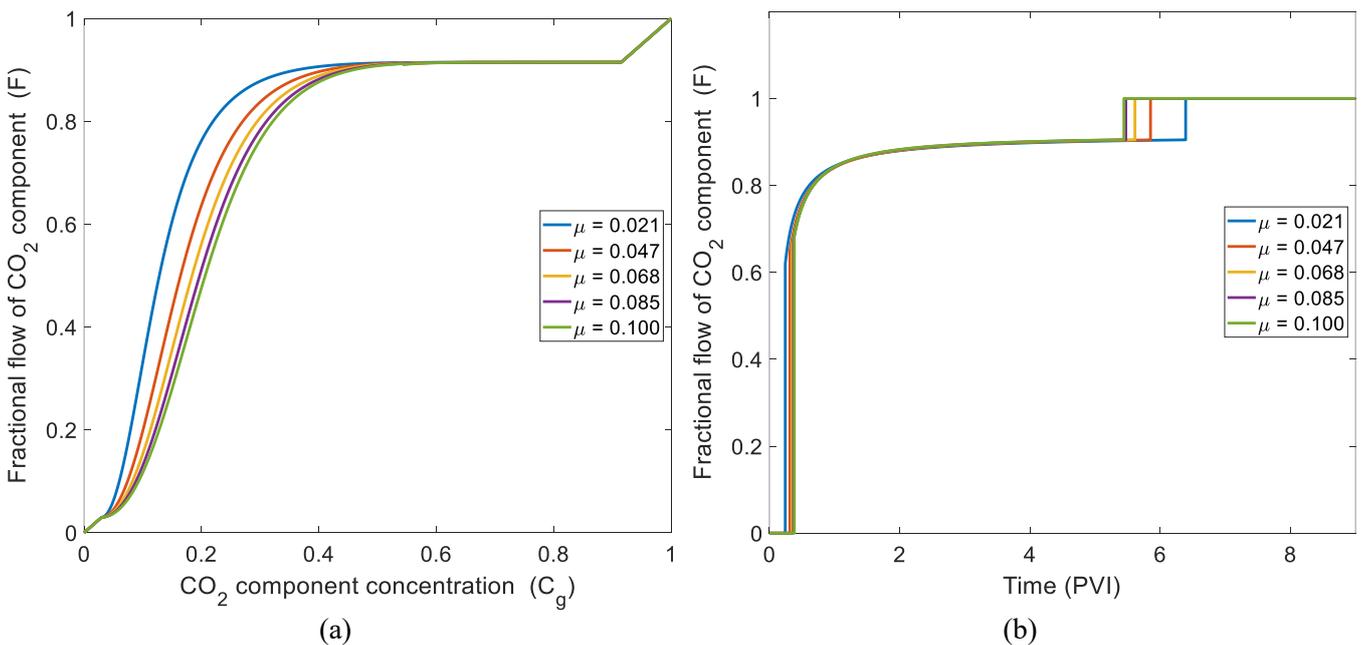

Fig. 5: (a) Fractional flow versus $CO_2$ concentration; (b) Fractional flow of $CO_2$ versus time at the outlet, $x=1$.

Fig. 5b shows the corresponding breakthrough gas-cuts $F(C_g)$ versus PVI at $x=1$. Blue curve is the breakthrough $CO_2$-cut, which is the $CO_2$-flux in both phases. After fast breakthrough of low-viscosity gas (at 0.2PVI), gas-cut jumps up to 0.65 and then quickly increasing until 0.9 at the moment of 6.4 PVI. Then gas-cut jumps at this moment to 1. Beyond this moment of long-term injection, the stabilisation and complete evaporation of water into the injected $CO_2$ occurs.

## 4.2   Pressure distribution in the reservoir

Let us derive an analytical model for the low-compressible flow ahead of the displacement front. Below we show that 2D single-phase low-compressible flow can be averaged in $z$, yielding traditional 1D pressure-diffusivity equation. The 2D mass balance equation for low-compressible water ahead of the advanced front is

$$\frac{\partial}{\partial t}(\phi \rho_w) + \frac{1}{r}\frac{\partial}{\partial r}(r \rho_w u_{wr}) + \frac{\partial}{\partial z}(\rho_w u_{wz}) = 0 \tag{35}$$

It is assumed that the rate at the advanced front is equal to the injection rate and the pressure at a location far from the injection well is equal to the reservoir pressure. Hence, the following boundary conditions are applied to Eq. (35)

$$\begin{cases} r = r_f(z_0), \, Q = Q_{inj} \\ p(\infty) = p_e \end{cases} \tag{36}$$

The gas and two-phase regions are considered incompressible. The introduction of the dimensionless parameter Eq. (A3) and the self-similar variable Eq. (32) to Eq. (36) results to

$$\begin{cases} \xi = D_0, \, Q = Q_{inj} \rightarrow \frac{\partial P}{\partial x} = -\frac{1}{\Lambda(S=0)x} \\ P(\infty) = P_e \end{cases} \tag{37}$$

The assumption of a low-compressible water phase yields the diffusivity equation Eq. (C5). Details of the derivation is given in Appendix C. By imposing the boundary conditions Eq. (37) on Eq. (C10), the pressure distribution ahead of the water-gas front accounting for low water compressibility is

$$P = -\frac{1}{\Lambda(0)} e^{(AD_0)} Ei(-A\xi) + P_e \tag{38}$$

## 4.3   Total mobility in two-phase region

The overall flux, as expressed in Eq. (A2), is the sum of the integrals of both fluxes presented in Eq. (A1) with respect to the depth ($z$) which yield the following dimensionless expression.

$$1 = -\Lambda(S)x\frac{\partial P}{\partial x} \tag{39}$$

where the total mobility is expressed as

$$\Lambda(S) = \left[ \int_0^1 \frac{K(Z)k_{rg}(s)}{\left[1+\beta\Delta\sigma_{cr}(s)\right]}dZ - \frac{\mu_g}{\mu_w}\int_0^1 K(Z)k_{rw}(s)dZ \right] \tag{40}$$

The detailed derivation is presented in Appendix D. The expression for total mobility for Zone 0, Zone 01 and Zone 1 are given by Eqs. (D5-D7) in Appendix D.

## 5   Derivations of expressions for sweep efficiency and well impedance

Based on the analytical model, presented in the previous section, the current section calculates pressure drawdown and well impedance (section 5.1) and sweep efficiency (section 5.2).

## 5.1 Pressure drop / well impedance: explicit formulae

Based on Eq. (39), the pressure drop at any given moment ($\tau$) is expressed as

$$\Delta P(\tau) = \int_{x_w}^{\infty} \left(-\frac{\partial P}{\partial x}\right) dx = \int_{x_w}^{\infty} \frac{1}{\Lambda(S)x} dx \qquad (41)$$

At any given moment $\tau_i < \tau(D_0)$, three distinct regions exist: the gas zone, the two-phase zone, and the water zone ahead of the $CO_2$ front as shown in. The pressure drop for this moment is expressed as

$$\Delta P(\tau) = \int_{x_w}^{D_1\tau} \frac{1}{\Lambda(1)x} dx + \int_{D_1\tau}^{D_0\tau} \frac{1}{\Lambda(S)x} dx + \int_{D_0\tau}^{\infty} \frac{dP}{dx} dx = I_1 + I_2 + I_3 \qquad (42)$$

where first, second and third integrals correspond to gas zone, mixture zone, and water zone, respectively.

At the two-phase zone the pressure drop is expressed via the self-similar variable, $\xi$, where

$$d\xi = F''(C_g) dC_g \qquad (43)$$

The introduction of the self-similar variable in Eq. (32) into the second term of Eq. (42) yields the pressure drop in the two-phase zone as

$$I_2 = \int_{C_{g1}}^{C_{g0}} \frac{F''(C_g)}{F'(C_g)\Lambda(C_g)} dC_g \qquad (44)$$

For the low-compressible water zone ahead of the advanced front, the pressure drop is calculated using the solution, Eq. (38) to the pressure diffusivity equation Eq. (C5) to obtain

$$I_3 = \int_{D_0\tau}^{\infty} -\frac{dP}{dx} dx = \left[P_e - \frac{1}{\Lambda(0)} e^{(AD_0)} Ei(-AD_0)\right] - [P_e] = -\frac{1}{\Lambda(0)} e^{(AD_0)} Ei(-AD_0) \qquad (45)$$

Substituting Eqs. (44-45) into Eq. (42) gives the pressure drop at moment $\tau$ as

$$\Delta P(\tau) = \int_{x_w}^{D_1\tau} \frac{1}{\Lambda(1)x} dx + \int_{C_{g1}}^{C_{g0}} \frac{F''(C_g)}{F'(C_g)\Lambda(C_g)} dC_g - \frac{1}{\Lambda(0)} e^{(AD_0)} Ei(-AD_0) \qquad (46)$$

Well impedance is the normalised reciprocal to injectivity index and can be expressed in terms of pressure drop as follows

$$J_w(\tau) = \frac{II(\tau=0)}{II(\tau)} = \frac{Q(\tau=0)}{\Delta p(\tau=0)} \frac{\Delta P(\tau)}{Q(\tau)} \qquad (47)$$

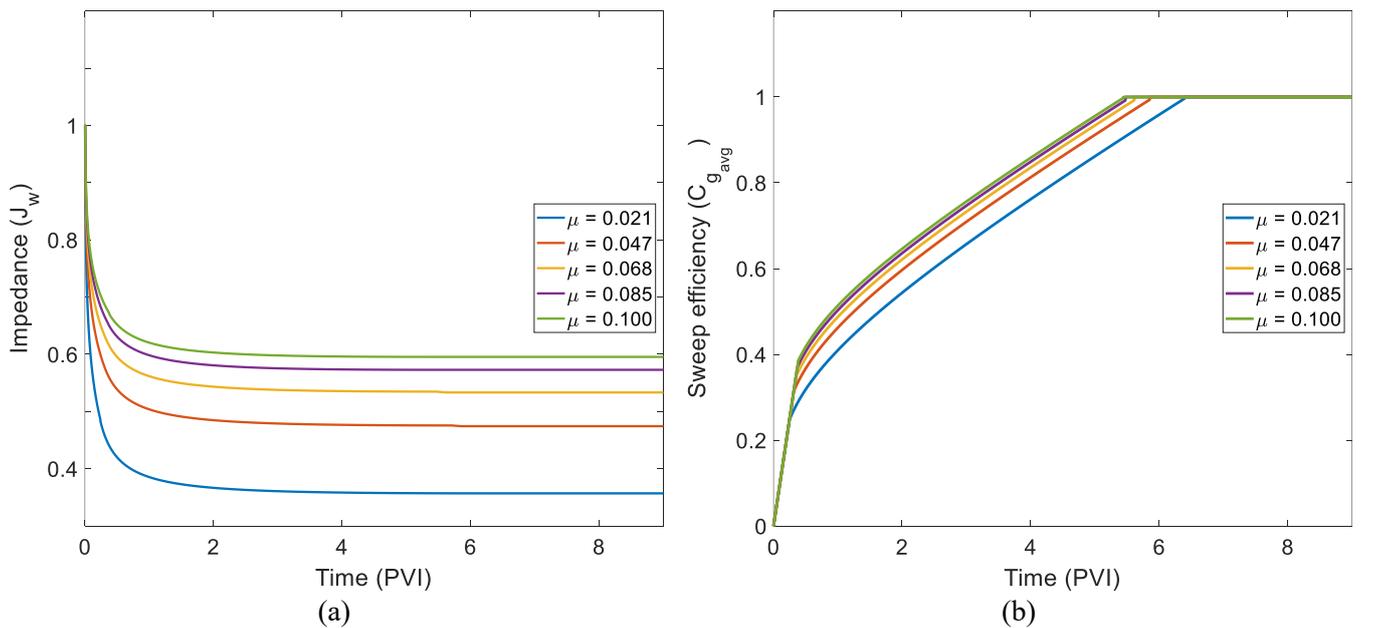

Fig. 6: (a) Impedance versus time; (b) Sweep efficiency versus time.

The results of impedance and sweep calculations are presented in Fig. 6. Here we consider layer-cake reservoir with $k(z)$ given by third-order polynomial

$$k(z) = (k_{max} - k_{min})\left(\frac{z}{H}\right)^3 + k_{min} \rightarrow k(Z) = (k_{max} - k_{min})Z^3 + k_{min} \tag{48}$$

The minimum and maximum permeabilities are $k_{min} = 2mD$, $k_{max} = 700mD$. In this study, we consider a reservoir of thickness, $H=25m$ with reservoir pressure of $P_e=400bars$ and temperature of $120^0C$. As it follows from phase diagram in Fig. 4b, $c_g = 0.03$, $c_w = 0.09$. The Leverette function for capillary pressure and Corey relative permeability are expressed as:

$$J(s) = C(s)^{-\eta}, \quad k_{rg}(s) = k_{rgwi}(s)^{n_g}, \quad k_{rg}(s) = k_{rwgr}(s)^{n_w}, \quad s = \frac{s^*}{1-S_{wi}} \tag{49}$$

where the Corey index, $\eta = 1.3$; the end point saturations of gas and water are 0 and 0.2, respectively; the end point relative permeability of gas and water are $k_{rgwi} = 0.6$ and $k_{rwgr} = 1$, The fluid exponents for gas and water are $n_g = 3$ and $n_w = 5$. The following typical values are used to describe formation damage due to fines migration, Eq. (6): power law exponent for straining, $B = 1.73$; total detachable fines concentration on the rock surface, $\sigma(l) = 7.12*10^{-4}$; the formation damage coefficient, $\beta = 1.5*10^{-4}$.

Blue curve in Fig. 6a shows the impedance decreases with time during the injection. Some increase in impedance due to fines migration is overly compensated by injection of gas with viscosity that is highly below the displaced water viscosity.

## 5.2 Derivations of sweep-efficiency expressions

The ODE Eq. (27) allows for first integral. Let us integrate both sides of Eq. (27) over the domain $\Omega$ in the plane $(x,\tau)$ limited by characteristics with velocity $D_0$ and $D_1$ as shown in Fig. 7. Calculating double integral of both sides of Eq. (27) over the domain $\Omega$ using Green's theorem yields:

$$\iint \left[\frac{\partial C_g}{\partial \tau} + \frac{\partial F}{\partial x}\right] d\tau dx = \oint Fd\tau - C_g dx = 0 \tag{50}$$

Accounting for the constancy of $CO_2$ concentration, $C_g$ along the characteristics, the contour integral in Eq. (50) yields

$$\overline{C_g}(\tau) = (1-x_w)C_g\left(\frac{1}{\tau}\right) - \tau F\left(C_g\left(\frac{1}{\tau}\right)\right) + \tau \tag{51}$$

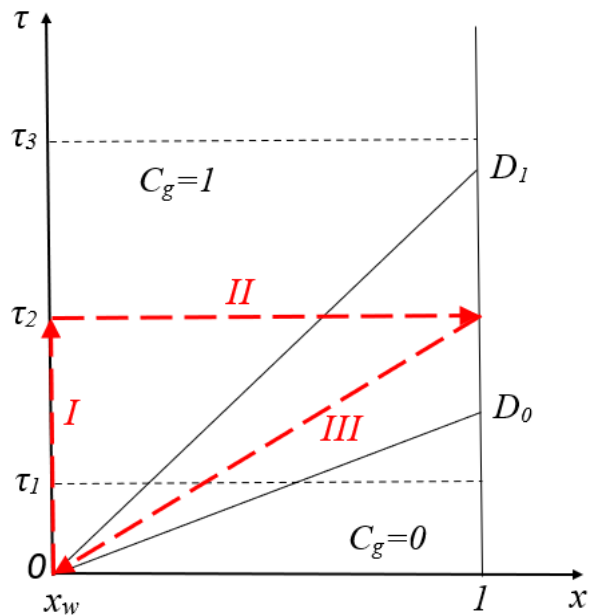

Fig. 7: Characteristic lines of $CO_2$ propagation in the plane $(x,\tau)$.

where the average CO₂ concentration is given as

$$\overline{C_g}(\tau) = \int_{x_w}^{1} C_g(x,\tau)dx \qquad (52)$$

At early moments in time, before the breakthrough of gas, $\left(0 < \tau < \dfrac{1}{D_0}\right)$, the average CO₂ concentration is given as

$$\overline{C_g}(\tau) = \tau \text{ with } C_g\left(\frac{1}{\tau}\right) = 0, \quad F\left(C_g\left(\frac{1}{\tau}\right)\right) = 0 \qquad (53)$$

At intermediate moments in time, $\left(\dfrac{1}{D_0} < \tau < \dfrac{1}{D_1}\right)$, the average CO₂ concentration is given as

$$\overline{C_g}(\tau) = (1-x_w)C_g\left(\frac{1}{\tau}\right) - \tau F\left(C_g\left(\frac{1}{\tau}\right)\right) + \tau \qquad (54)$$

At late moments, when stabilization is established $\left(\dfrac{1}{D_1} < \tau < \infty\right)$, the average CO₂ concentration is given as

$$\overline{C_g}(\tau) = (1-x_w) \text{ with } C_g\left(\frac{1}{\tau}\right) = 1, \quad F\left(C_g\left(\frac{1}{\tau}\right)\right) = 1 \qquad (55)$$

Blue curve in Fig. 6b shows linear sweep growth before the breakthrough at the moment 0.2PVI, following slower increase during CO₂ production in two phases. Full evaporation occurs at the moment 6.4 PVI.

## 6 Methodology of downscaling

The formulae Eqs. (16-17) of phase and component distributions with depth versus average gas concentrations allow to downscale, i.e. to restore depth distributions from the average concentration $C_g(x,\tau)$. This section uses the analytical model for upscaled flow Eq. (27) to determine depth saturation distributions.

Fig. 8a shows the case of permeability increasing with depth, Fig. 8b – depth decreasing permeability. Viscous forces direct the injected gas into highly permeable layers. Buoyancy forces gas upwards. In the case of permeability increasing with depth, both viscous and gravitational forces move gas upwards. So, gas saturation decrease with depth is expected. Indeed, in Fig. 9a, where saturation variation with depth is presented for 7 moments, gas saturation decreases with depth.

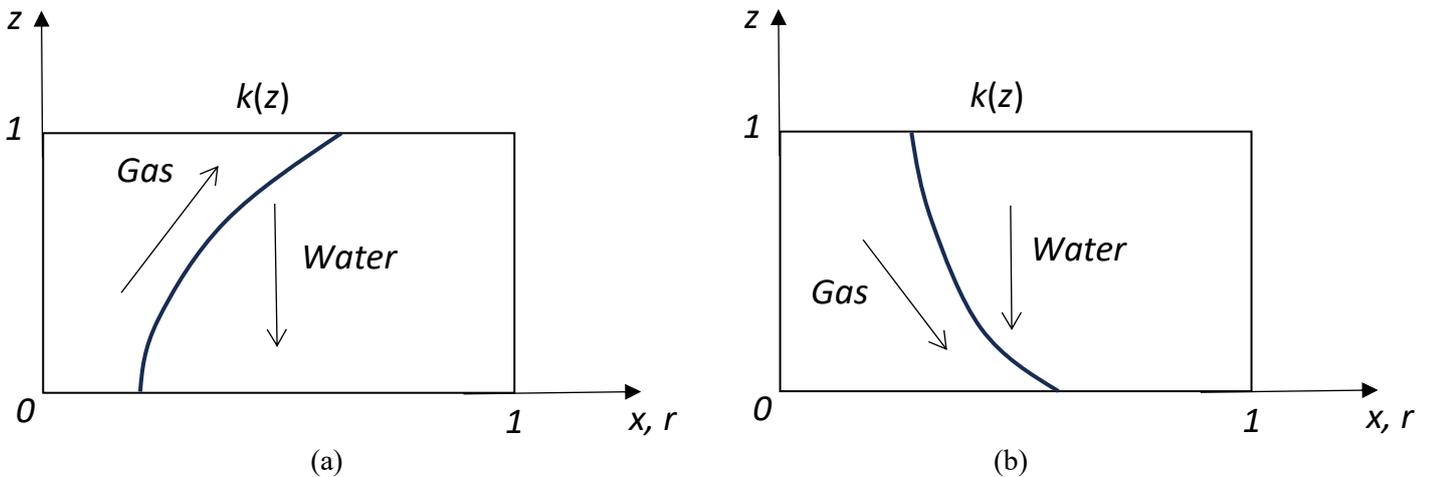

*Fig. 8: Schematic representation of flow regimes for a layer cake reservoir: (a) Permeability decreasing with depth; (b) Permeability increasing with depth.*

In cases where permeability decreases with depth, viscous forces tend to direct the injected gas into the lower reservoir layers, while gravitational forces drive the gas upward. The interplay between these opposing forces results in non-monotonic vertical saturation profiles. As illustrated in Fig. 9b gas saturation increases from the reservoir bottom upward, reaching a maximum at an intermediate depth, and subsequently decreases toward the reservoir top. At each depth, saturation monotonically increases with time.

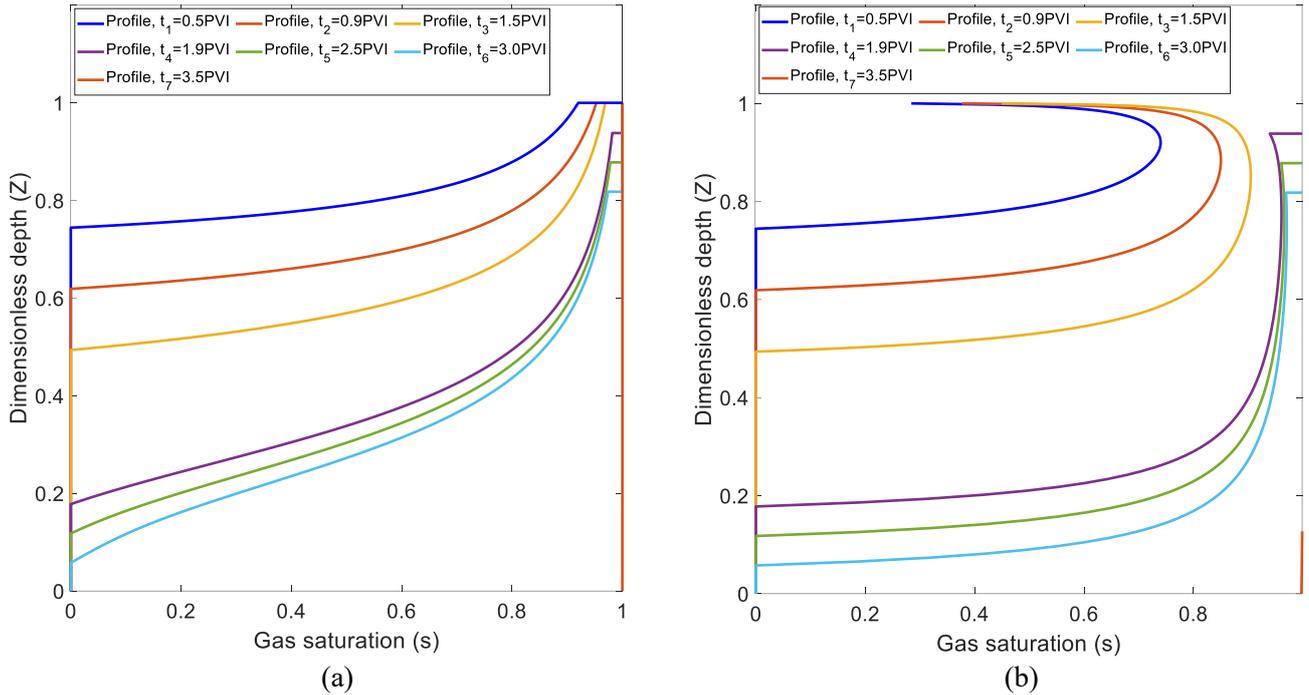

*Fig. 9: Vertical saturation distributions obtained from model downscaling at various times for:* (*a*) *permeability decreasing with depth;* (*b*) *permeability increasing with depth.*

## 7  Multivariant sensitivity analysis

This section presents sensitivity study of water displacement by injected $CO_2$ versus reservoir heterogeneity, viscosity ratio, concentration of detachable fines, formation damage coefficient and power law exponent for straining.

*Heterogeneity* – Here we use the relative range, ν, that is equal to the difference between maximum and minimum permeability divided by its mean value. The relative range expresses the spread of permeability values relative to the average; a higher relative range indicates higher heterogeneity, while a lower relative range implies more uniform permeability. Fig. 10a shows that the more heterogeneous is the reservoir, the higher is the fractional flow and the more concave is the fractional flow curve. Consequently, the higher is the gas cut (Fig. 10b). Fig. 10(b-d) show that increase in heterogeneity decrease breakthrough period from 0.4PVI to 0.2PVI, increase the evaporation period from 3.5PVI to 4.5PVI, decrease sweep by 35% and decrease the stabilized impedance from 0.45 to 0.36.

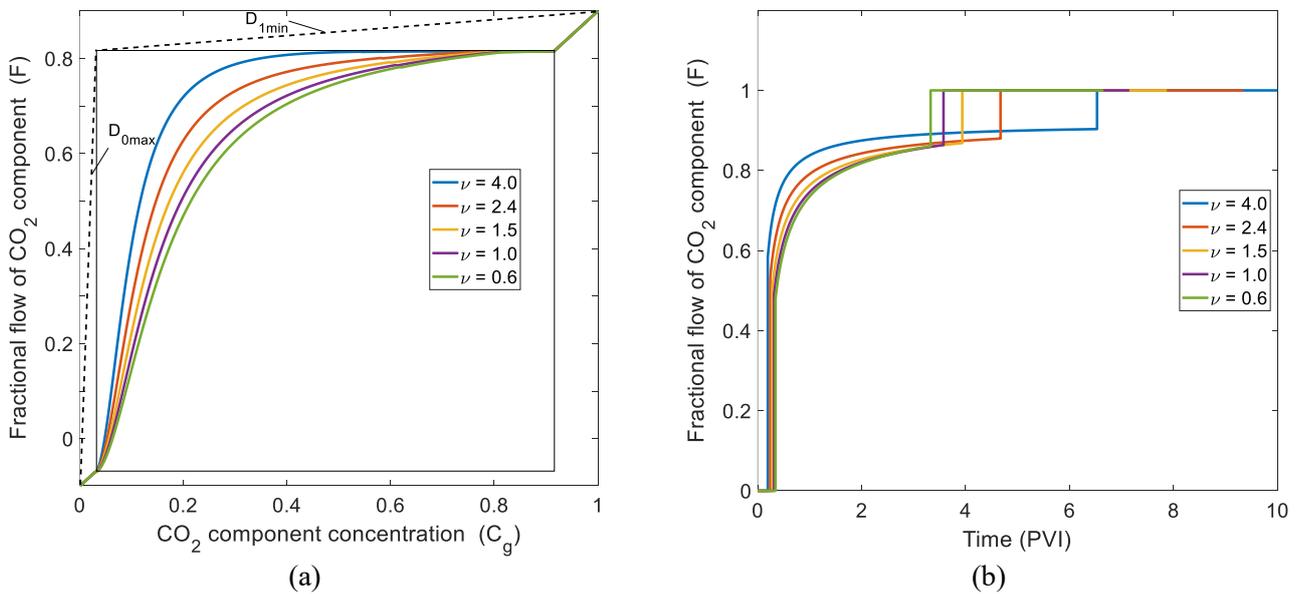

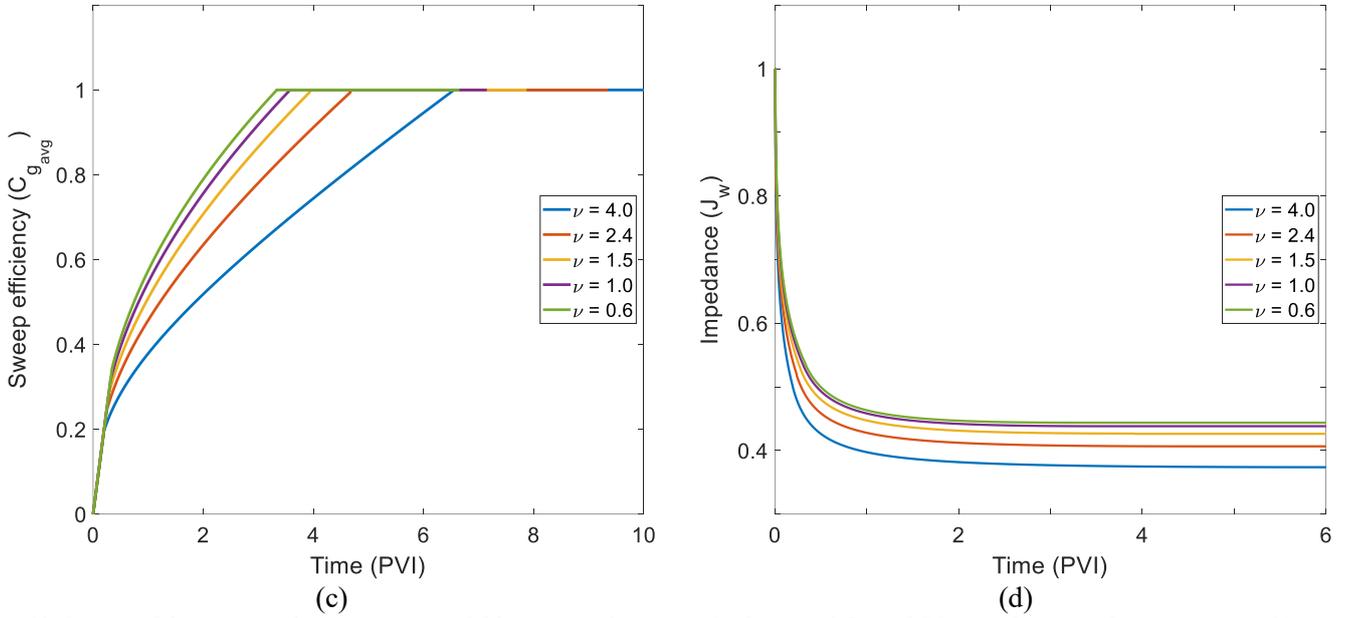

Fig. 10: Impact of the reservoir heterogeneity on CO2 injection dynamics: (a) fractional flow of $CO_2$ as a function of concentration, (b) outlet fractional flow evolution over time, (c) sweep efficiency over time, and (d) impedance variation with time.

The curvilinearity increase (Fig. 10a) results to the decrease of tangent $D_1$ and increase in tangent $D_0$. So, increase in heterogeneity decreases breakthrough time, $1/D_0$ and increases the evaporation period, $1/D_1$, as shown in Fig. 10b and Fig. 10c. However, even for very concave FFC, $D_1$ value is lower than the tangent of the straight-line segment 1-L in Fig. 4b; $D_0$ cannot exceed the tangent of the straight-line segment 0-L

$$D_0 > \frac{1-c_w}{c_g}, \quad D_1 < \frac{c_w}{1-c_g} \qquad (56)$$

Consequently, breakthrough time is lower than $c_g/(1-c_w)$, and the evaporation period is above $(1-c_g)/c_w$. The higher is the heterogeneity, the lower is the sweep (Fig. 10c) and the lower is the impedance (Fig. 10d).

*Viscosity ratio* –Fig. 5a presents the plots of fractional flow functions $F(C_g)$ for various viscosity ratios $\mu=\mu_g/\mu_w$. Here the viscosity ratio increases from 0.021 typical for gaseous $CO_2$ up to 0.1 for supercritical $CO_2$. The higher is the gas viscosity, the lower is its fractional flow. The single-phase regions for fractional flow functions are independent of phase viscosities. Fig. 5b shows the corresponding breakthrough gas-cuts ($F(C_g)$) versus PVI at $x = 1$. The higher is the gas viscosity, the higher is its fraction at the outlet. Consequently, the higher is the viscosity ratio, the higher is the sweep efficiency (Fig. 6b). With the overall viscosity ratio increase, the sweep efficiency of the heterogeneous reservoir increases by 0.1, which is quite significant. Simultaneously, the stabilised well impedance increases approximately twice (Fig. 6a).

*Concentration of detachable fines* –Fig. 11 shows the effects of total concentration of detachable fines. The higher is the detachable fines concentration, the higher is the concentration of strained fines and the consequent formation damage to gas phase and gas deceleration. Fig. 11a shows that increase in $\sigma(1)$ yields increase in breakthrough time and decrease in evaporation period. It also yields the increase in sweep by 0.14 with the overall $\sigma(1)$ increase from $5*10^{-5}$ to $4*10^{-3}$, (Fig. 11b). Overall increase in detachable fines concentration results in impedance increase; the stabilized values increase 1.75 times.

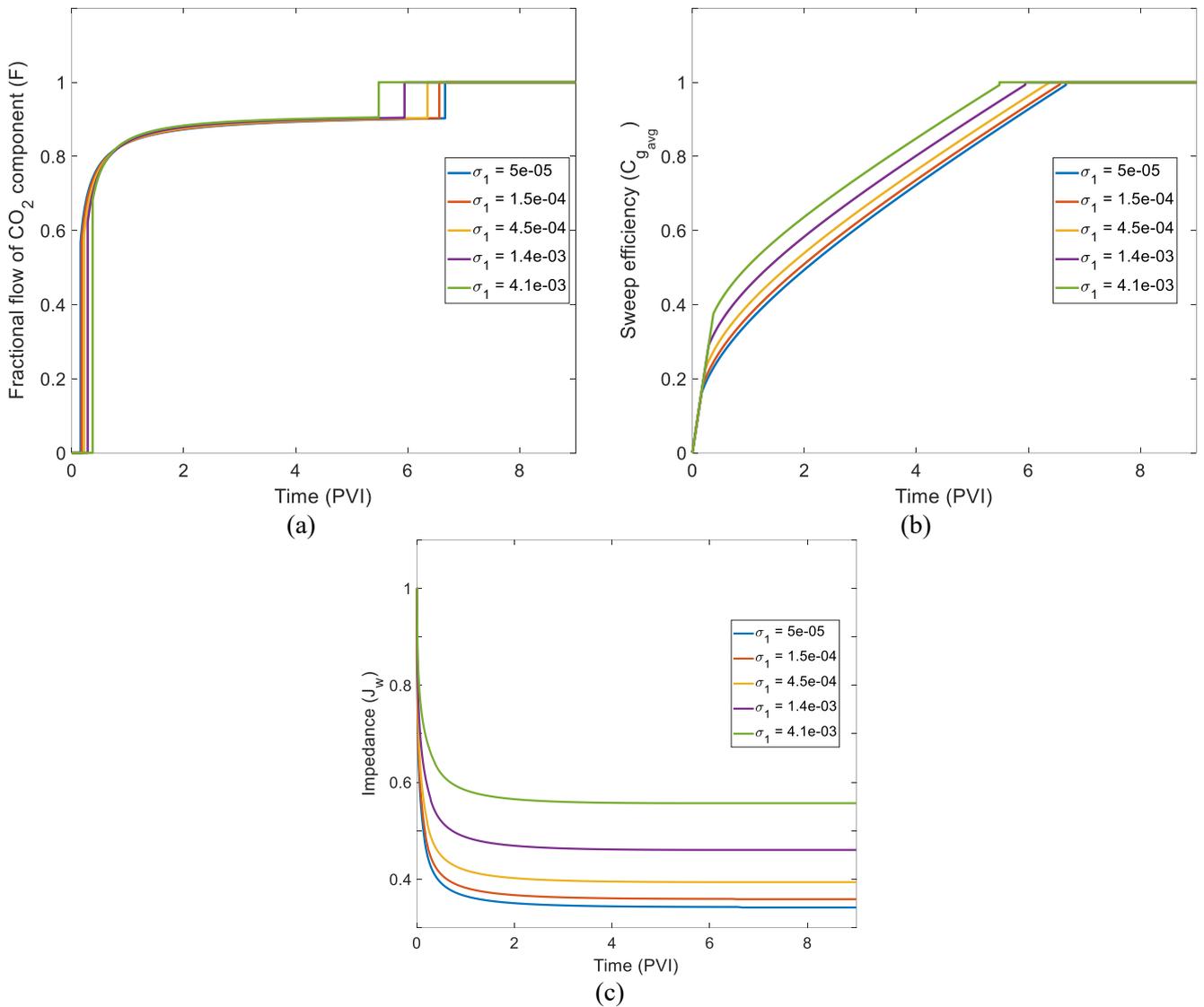

*Fig. 11: Impact of the concentration of detachable fines on CO$_2$ injection dynamics: (a) fractional flow of CO$_2$ as a function of concentration, (b) outlet fractional flow evolution over time, (c) sweep efficiency over time, and (d) impedance variation with time.*

*Formation damage coefficient* – The concentration of detachable fines $\sigma(1)$ and formation damage coefficient $\beta$ are expressed as product in Eq. (6). As a result, increasing $\sigma(1)$ has the same effect as increasing $\beta$; for example, a hundredfold increase in $\sigma(1)$ is equivalent to a hundredfold increase in $\beta$. Fig. 12a demonstrates that increasing the formation damage coefficient from zero to 30,000 extends the breakthrough period from 0.2 PVI to 0.4 PVI, and reduces the evaporation period from 6.8 PVI to 5.8 PVI. Additionally, as shown in Fig. 12b, this increase in $\beta$ leads to 30% improvement in sweep efficiency and raises the stabilized impedance from 0.34 to 0.58.

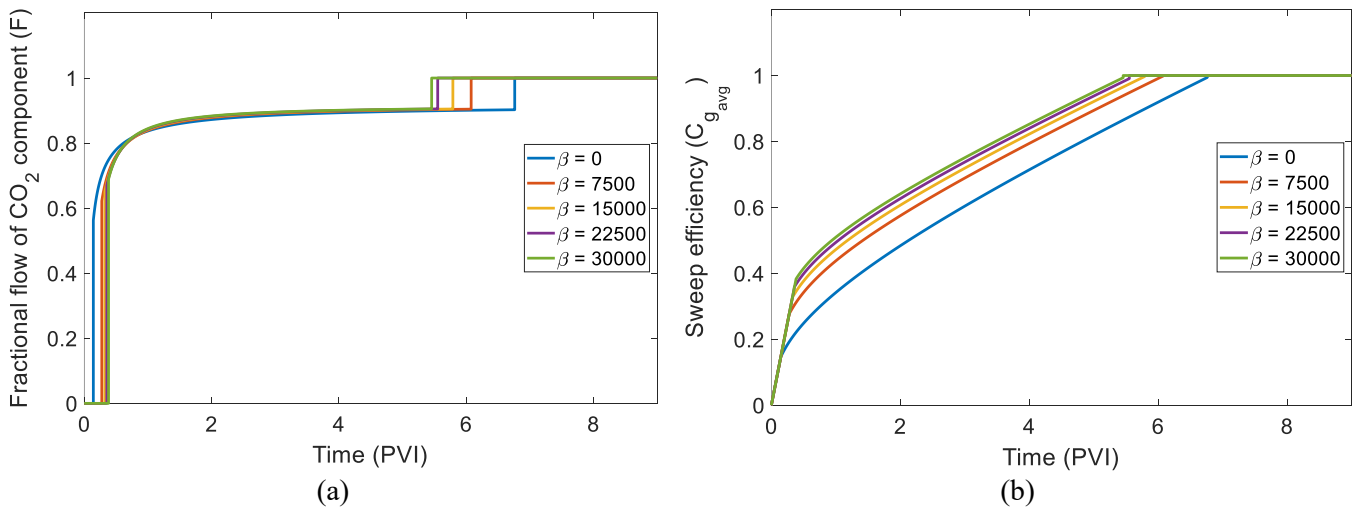

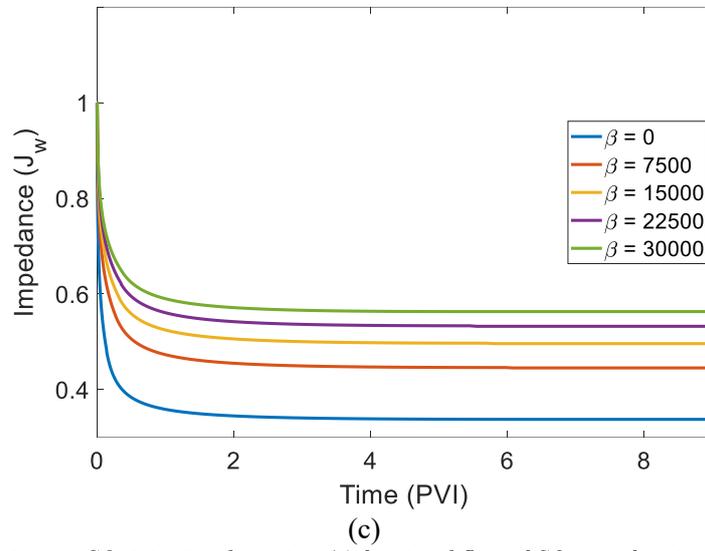

*Fig. 12: Impact of damage coefficient on $CO_2$ injection dynamics: (a) fractional flow of $CO_2$ as a function of concentration, (b) outlet fractional flow evolution over time, (c) sweep efficiency over time, and (d) impedance variation with time.*

*Power law exponent for strained fines* – Eq. (6) shows that the higher is the power law exponent *B*, the higher is the MRF, and the higher is the damage to gas phase. Fig. 13 shows that increase of power law *B* yields increase in sweep efficiency by 11% (Fig. 13b), increase in breakthrough period and decrease in overall evaporation period (Fig. 13a) and increase in stabilized impedance from 0.45 to 0.52 (Fig. 13c).

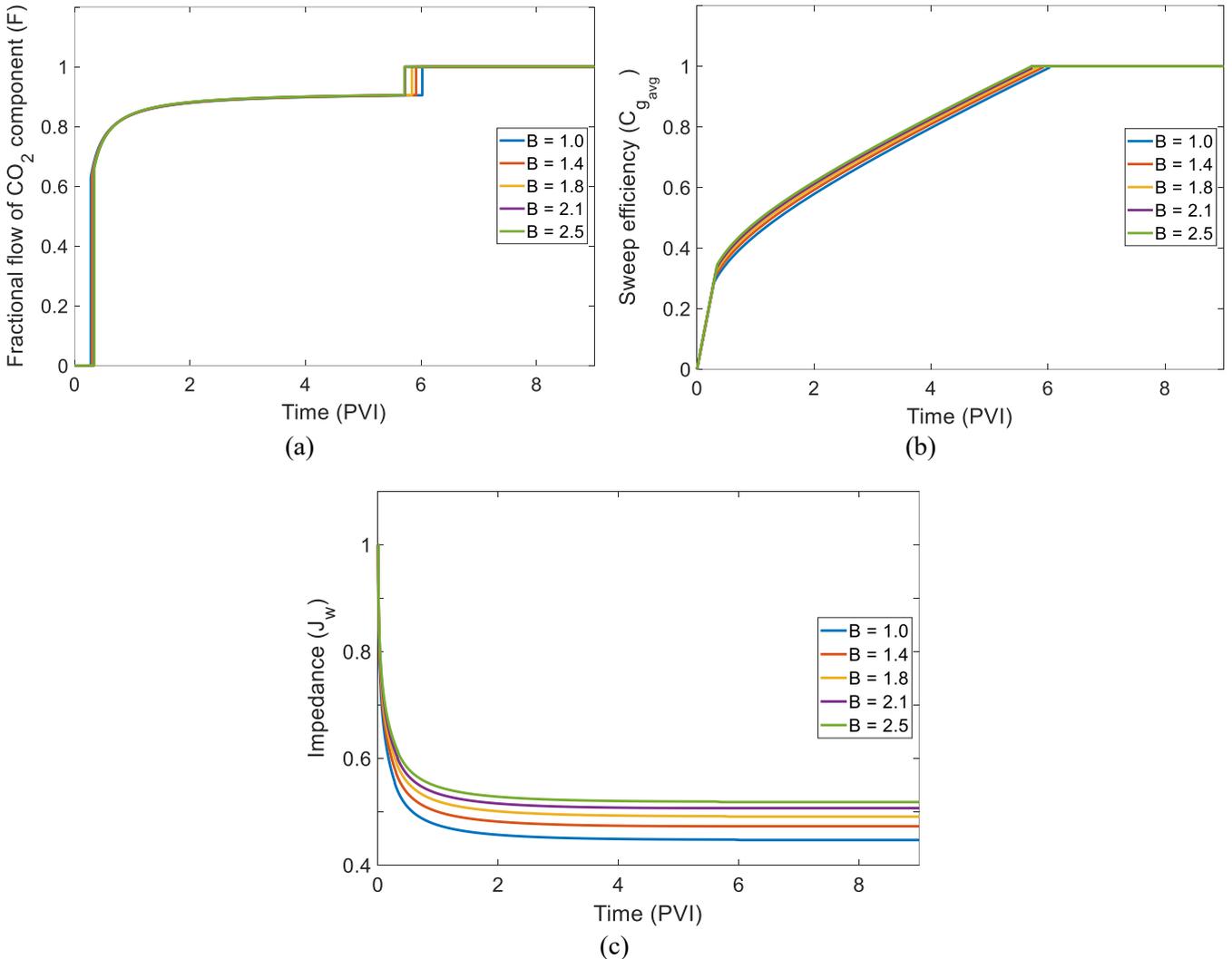

*Fig. 13: Impact of the power law exponent for straining on $CO_2$ injection dynamics: (a) fractional flow of $CO_2$ as a function of concentration, (b) outlet fractional flow evolution over time, (c) sweep efficiency over time, and (d) impedance variation with time.*

## 8 Summary and discussions

Let us discuss the restrictions of the developed analytical model for predicting of $CO_2$ storage in saline aquifers.

The conditions under which the VE model is valid have been well established both theoretically and through validation against two-dimensional (2D) and three-dimensional (3D) reservoir simulations [Igor Yortsos Shapiro]. In the context of immiscible fluid displacement in layered reservoirs, the model assumes vertical equilibrium, which has been widely validated under typical reservoir conditions. The assumption of the maximum retention function used to model the detachment of fines was derived through exact upscaling from pore-scale to core-scale [48]. Additionally, the assumption that straining of mobilised fines occurs instantaneously, coupled with the thermodynamic equilibrium between $CO_2$ and water, is supported by the fact that the corresponding relaxation times are significantly smaller than the time required for one pore volume injection (PVI). Thus, the extension of the VE model to account for partial $CO_2$-water miscibility and fines migration does not introduce additional restrictions beyond those imposed by the traditional VE framework.

Laboratory core flooding experiments, involving $CO_2$ injection into brine-saturated cores, typically exhibit between $0.5*10^4$ and $2.0*10^4$ PVIs during the full evaporation period [47]. In contrast, the large-scale VE model presented in Eq. (31) describes the full evaporation period as $1/D_1$, where the period typically varies within the range of $[10, 10^3]$. The discrepancy between these scales arises from the differing nature of the two processes: core flooding experiments are conducted under non-equilibrium evaporation conditions, while large-scale reservoir displacement follows the assumptions of the large-scale approximation, as outlined by previous work [28]. The large-scale VE model, Eq. (31) does not account for capillary pressure effects, which could influence fluid distribution within the reservoir. With capillary pressure, Eq. (25) allows for exact self-similar solution, where initial saturation is achieved asymptotically as the radius approaches infinity. Obtaining this solution is the topic of forthcoming research.

Several studies by F. Hussain and collaborators have investigated the effects of $CO_2$ humidity on reservoir sweep efficiency and plume propagation, particularly in the context of $CO_2$ injection with equilibrium vapor concentrations. To assess the impact of $CO_2$ humidity on the displacement behaviour in layer-cake reservoirs, the current VE model (Eq. 31) would need to be extended to account for injected $CO_2$ with vapor content. This extension is crucial for improving the model's predictive capability in reservoirs where $CO_2$ injection with significant vapor content may alter the displacement dynamics.

## 9 Conclusions

Analytical modelling of radial, pseudo-2D displacement of brine by $CO_2$, accounting for reservoir heterogeneity, partial water-gas miscibility, and fines migration, leads to the following key conclusions.

The vertical capillary-gravity equilibrium (VE) model for two-phase flow in layer-cake reservoirs can be extended to incorporate water evaporation into the injected gas, $CO_2$ dissolution in formation brine, and fines migration with associated permeability damage. The extended VE formulation for unknown averaged $CO_2$ concentration $C_g(r,t)$ is mathematically equivalent to a 1D model of partially miscible displacement of binary mixtures with mobile fines in a homogeneous formation. The VE model for unknown pressure $p(r,t)$ for incompressible flow can be matched to the initial reservoir pressure by introducing slight compressibility in the reservoir fluid ahead of the displacement front, enabling continuity of pressure across regions.

The quasi-2D vertically averaged model allows for an exact self-similar solution, yielding explicit formulae for phase saturations, $CO_2$ concentration, and pressure. This analytical solution also provides explicit formulae for estimating sweep efficiency (storage capacity) and well injectivity, enabling rapid multivariant sensitivity study and parametric evaluations.

Downscaling of the 1D solution reconstructs vertical ($z$-dependent) profiles of saturation and concentration as functions of radial position and time. This is achieved through parameterization of depth-dependent distributions based on the averaged flow variables. Downscaling of vertical $CO_2$-concentration and saturation distribution in the reservoirs with the permeability decreasing with depth yields monotonic $CO_2$-concentration and saturation distributions. If the permeability increases with depth, $CO_2$-concentration and saturation distribution can be nonmonotonic, which is attributed to the force competition between gravity and pressure gradient.

Sensitivity analysis reveals that $CO_2$-brine viscosity ratio, permeability heterogeneity, and fines migration significantly influence storage performance and injectivity. Increased permeability damage reduces injectivity but enhances storage capacity due to improved sweep.

Higher permeability heterogeneity leads to earlier gas breakthrough and a longer evaporation period. However, for any depth-dependent permeability distribution $k(z)$, breakthrough time remains below $c_g/(1-c_w)$, while the duration of the evaporation period exceeds $(1-c_g)/c_w$.

**Appendix A:** Expressions for overall gas and water fluxes

The overall phase fluxes of water and gas in the radial direction are obtained by integration of Eq. (8) over reservoir thickness

$$U_{wr}^* = \int_0^H u_{wr} dz = -\int_0^H \frac{k(z)k_{rw}(s)dz}{\mu_w} \frac{\partial p_w}{\partial r}, \quad U_{gr}^* = \int_0^H u_{gr} dz = -\int_0^H \frac{k(z)k_{rg}(s)dz}{\mu_g[1+\beta\Delta\sigma_{cr}(s)]} \frac{\partial p_g}{\partial r} \quad (A1)$$

Taking the sum of the phase velocities in Eq. (A1) and expressing the result in terms of pressure in water and the capillary pressure yields

$$U = -\left[\int_0^H \frac{k(z)k_{rw}(s)dz}{\mu_w} + \int_0^H \frac{k(z)k_{rg}(s)dz}{\mu_g[1+\beta\Delta\sigma_{cr}(s)]}\right]\frac{\partial p_w}{\partial r} - \int_0^H \frac{k(z)k_{rg}(s)dz}{\mu_g[1+\beta\Delta\sigma_{cr}(s)]}\frac{\partial p_c(s)}{\partial r} \quad (A2)$$

Let us introduce dimensionless variables and parameters:

$$x=\frac{r^2}{R^2}, \; Z=\frac{z}{H}, \; \tau=\frac{Qt}{\phi\pi R^2}, \; h_c=\frac{H_c}{H}, \; \delta=\frac{4\pi k_0 H}{Q\mu_w}\frac{\sigma\cos\theta\sqrt{\phi}}{\sqrt{k_0}}, \; \kappa=\frac{4\pi k_0}{Qc_t\mu_w}, \; p=\frac{1}{H}\int_0^H \tilde{p}(x,z,t)dz$$

$$P=\frac{4\pi k_0 H}{Q\mu_w}p, \; P_w=\frac{4\pi k_0 H}{Q\mu_w}p_w, \; k_0=\frac{1}{H}\int_0^H k(z)dz = \int_0^1 k(Z)dZ, \; k(Z)=k_0 K(Z), \; \int_0^1 K(Z)dZ=1$$

(A3)

Based on the dimensionless variables and parameters in Eq. (A3), the dimensionless form of the overall phase flux for water and gas is obtained as:

$$1 = -\left[\int_0^1 K(Z)k_{rw}(s)dZ + \int_0^1 \frac{K(Z)k_{rg}(s)dZ}{\mu[1+\beta\Delta\sigma_{cr}(s)]}\right]x\frac{\partial P_w}{\partial x} - \delta\int_0^1 \frac{K(Z)k_{rg}(s)dZ}{\mu[1+\beta\Delta\sigma_{cr}(s)]}x\frac{\partial J(s)}{\partial x}$$

Substituting Eq. (A1) into the second equation of Eq. (2) yields the overall flux of $CO_2$-component as

$$U_g = -(1-c_w)\int_0^H \frac{k(z)k_{rg}(s)dz}{\mu_g[1+\beta\Delta\sigma_{cr}(s)]}\frac{\partial p_g}{\partial r} - c_g\int_0^H \frac{k(z)k_{rw}(s)dz}{\mu_w}\frac{\partial p_w}{\partial r} \quad (A4)$$

Expressing Eq. (A4) in terms of water and capillary pressure gradients gives

$$U_g = -\left[(1-c_w)\int_0^H \frac{k(z)k_{rg}(s)dz}{\mu_g[1+\beta\Delta\sigma_{cr}(s)]} + c_g\int_0^H \frac{k(z)k_{rw}(s)dz}{\mu_w}\right]\frac{\partial p_w}{\partial r} - (1-c_w)\int_0^H \frac{k(z)k_{rg}(s)dz}{\mu_g[1+\beta\Delta\sigma_{cr}(s)]}\frac{\partial p_c(s)}{\partial r} \quad (A5)$$

Determining the expression for water pressure gradient from Eq. (A2) and substituting the results into Eq. (A5) gives the overall flux of $CO_2$ component as

$$U_g = \frac{(1-c_w)\lambda_g(s)+c_g\lambda_w(s)}{\lambda(s)}U + \left\{\frac{(1-c_w)\lambda_g(s)+c_g\lambda_w(s)}{\lambda(s)}\lambda_g(s)-(1-c_w)\lambda_g(s)\right\}\frac{\partial p_c(s)}{\partial r} \quad (A6)$$

where the mobility of gas, water and total mobility are expressed as

$$\lambda_g(s) = \int_0^H \frac{k(z)k_{rg}(s)dz}{\mu_g[1+\beta\Delta\sigma_{cr}(s)]}, \quad \lambda_w(s) = \int_0^H \frac{k(z)k_{rw}(s)dz}{\mu_w}, \quad \lambda(s) = \lambda_g(s)+\lambda_w(s) \quad (A7)$$

Eqs. (A6, A7) are used in sections 3.1 and 3.2 to derive averaged mass balance equation (25) in three flow regions of thin and thick reservoirs.

**Appendix B:** 1D Averaged mass balance for gas

The 1D mass balance for displacement of water by gas through a layer cake system under vertical equilibrium is expressed as

$$\phi \frac{\partial C_g}{\partial t} + \frac{1}{r}\frac{\partial (rU_g)}{\partial r} = 0 \tag{B1}$$

Substituting the expression for overall $CO_2$ gas flux Eq. (24) into the mass balance Eq. (B1) gives

$$\phi \frac{\partial C_g}{\partial t} + \frac{1}{r}\frac{\partial}{\partial r}(rUF) = -\frac{1}{r}\frac{\partial}{\partial r}\left[r\left(F-(1-c_w)\right)\int_0^H \frac{k(z)k_{rg}(s)dz}{\mu_g(1+\beta\Delta\sigma_{cr}(s))}\frac{\partial p_c(s)}{\partial r}\right] \tag{B2}$$

Introducing the dimensionless variables and parameters (A3) into Eq. (B2) results in Eq. (25) presented in the main text.

**Appendix C:** Pressure distribution ahead of the water-gas front accounting for water compressibility

Let us show that 2D single-phase low-compressible flow under vertical gravity equilibrium can be averaged in $z$, yielding traditional 1D pressure-diffusivity equation. The 2D mass balance equation for low-compressible water ahead of the advanced front is

$$\frac{\partial}{\partial t}(\phi\rho_w) + \frac{1}{r}\frac{\partial}{\partial r}(r\rho_w u_{wr}) + \frac{\partial}{\partial z}(\rho_w u_{wz}) = 0 \tag{C1}$$

Under the assumption of vertical equilibrium (second Eq. (3)), the velocity of water in the vertical direction $u_{wz}$ is equal zero. Hence the mass balance equation for low-compressible water, Eq. (C1) reduces to

$$\frac{\partial}{\partial t}(\phi\rho_w) + \frac{1}{r}\frac{\partial}{\partial r}(r\rho_w u_{wr}) = 0 \tag{C2}$$

Substitution of the Darcy law for water, Eq. (8) and Eq. (9) for the total compressibility into Eq. (C2) and neglecting pressure dependency of water density in the mass transfer term yields

$$c_t \phi_0 \frac{\partial p}{\partial t} = \frac{1}{r}\frac{\partial}{\partial r}\left(r\frac{k(z)}{\mu_w}\frac{\partial \tilde{p}}{\partial r}\right) \tag{C3}$$

where $z$ becomes a parameter. Flows of water in all layers are independent, i.e. there is communication of between the layers. Let us consider the solution $p(r,0,t)$ at $z = 0$.
Eq. (C3) becomes

$$\frac{\partial p}{\partial t} = \frac{1}{Hc_t\phi_0\mu_w}\frac{1}{r}\frac{\partial}{\partial r}\left(r\int_0^H k(z)dz \frac{\partial p}{\partial r}\right) \tag{C4}$$

The solution of quasi 2D problem is obtained by extending $p(r, z = 0, t)$ to $0<z<1$ by hydrostatic equation.

*Solution of outflow problem*   Introducing the dimensionless variables and parameters, including the expression for average pressure, $p$, Eq. (A3) into Eq. (C4) yields the dimensionless form of the pressure diffusivity as

$$\frac{\partial}{\partial x}\left(x\frac{\partial P}{\partial x}\right) = A\frac{\partial P}{\partial \tau}, \quad A = 1/\kappa \tag{C5}$$

The introduction of the self-similar variable in Eq. (32) into Eq. (C5) yields a second order ordinary differential equation given as

$$\frac{dP}{d\xi} + \xi\frac{d^2P}{d\xi^2} = -A\xi\frac{dP}{d\xi} \tag{C6}$$

The order of Eq. (C6) is reduced by introducing $\omega = \dfrac{dP}{d\xi}$ to obtain

$$\omega(1 + A\xi) = -\xi \frac{d\omega}{d\xi} \tag{C7}$$

By the method of separation of variables, both sides of Eq. (C7) are integrated to obtain

$$-\ln(\omega) = \ln(\xi) + A\xi + c_1 \quad \Rightarrow \quad \frac{dP}{d\xi} = \frac{c_1}{\xi} e^{(-A\xi)} \tag{C8}$$

The solution to the ordinary differential equation (ODE), Eq. (C8) is

$$P = c_1 Ei(-A\xi) + c_2 \tag{C9}$$

where $c_1$ and $c_2$ are constants of integration

The boundary conditions, Eq. (37) are introduced to determine $c_1$ and $c_2$, which yield the particular solution for the diffusivity Eq. (C5) as

$$P = -\frac{1}{\Lambda(0)} e^{(AD_0)} Ei(-A\xi) + P_e \tag{C10}$$

**Appendix D:** Total mobility in two phase region and impedance calculation

The sum of the overall gas and water velocities, Eq. (A1) results to

$$2\pi r U = Q = -\left[\int_0^H \frac{k(z)k_{rw}(s)dz}{\mu_w} + \int_0^H \frac{k(z)k_{rg}(s)dz}{\mu_g \left[1 + \beta\Delta\sigma_{cr}(s)\right]}\right] 2\pi r \frac{\partial p_w}{\partial r} - \int_0^H \frac{k(z)k_{rg}(s)dz}{\mu_g \left[1 + \beta\Delta\sigma_{cr}(s)\right]} 2\pi r \frac{\partial p_c(s)}{\partial r} \tag{D1}$$

Substituting the dimensionless parameters Eq. (A3) into Eq. (D1) gives

$$1 = -\left[\int_0^1 K(Z)k_{rw}(s)dZ + \int_0^1 \frac{K(Z)k_{rg}(s)dZ}{\mu\left[1 + \beta\Delta\sigma_{cr}(s)\right]}\right] x \frac{\partial P_w}{\partial x} - \delta \int_0^1 \frac{K(Z)k_{rg}(s)dZ}{\mu\left[1 + \beta\Delta\sigma_{cr}(s)\right]} x \frac{\partial J(s)}{\partial x} \tag{D2}$$

At large-scale approximations, $\delta \ll 1$, hence the pressure drop in gas and water are equal, that is,

$$\frac{\partial P_g}{\partial x} = \frac{\partial P_w}{\partial x} = \frac{\partial P}{\partial x} \tag{D3}$$

Considering Eq. (D2), the expression for total flux becomes

$$1 = -\left[\int_0^1 K(Z)k_{rw}(s)dZ + \int_0^1 \frac{K(Z)k_{rg}(s)dZ}{\mu\left[1 + \beta\Delta\sigma_{cr}(s)\right]}\right] x \frac{\partial P}{\partial x} \tag{D4}$$

The total mobility for all the three zones (0, 01, and 1) defined by Eq. (40) are determined by the following expression.

For zone 0,

$$\Lambda_0(s) = \int_{Z_0}^1 \frac{K(Z)k_{rg}(s)dZ}{\mu\left[1 + \beta\Delta\sigma_{cr}(s)\right]} + \int_0^{Z_0} K(Z)dZ \tag{D5}$$

For zone 01,

$$\Lambda_{01}(s) = \int_0^{Z_0} K(Z)dZ + \int_{Z_0}^{Z_0+h_c} k_{rw}(s)K(Z)dZ + \int_{Z_0}^{Z_0+h_c} \frac{K(Z)k_{rg}(s)dZ}{\mu\left[1 + \beta\Delta\sigma_{cr}(s)\right]} + \frac{k_{rgwc}}{\mu\left[1 + \beta\Delta\sigma_{cr}\right]} \int_{Z_0+h_c}^1 k(Z)dZ \tag{D6}$$

For zone 1,

$$\Lambda_1(s) = \frac{k_{rgwc}}{\mu\left[1 + \beta\Delta\sigma_{cr}\right]} \int_{Z_0+h_c}^1 k(Z)dZ + \int_0^{Z_1} \left(k_{rw}(s)K(Z) + \frac{K(Z)k_{rg}(s)}{\mu\left[1 + \beta\Delta\sigma_{cr}(s)\right]}\right)dZ \tag{D7}$$